\documentclass{sig-alternate}

\pdfoutput = 0

\newfont{\mycrnotice}{ptmr8t at 7pt}
\newfont{\myconfname}{ptmri8t at 7pt}
\let\crnotice\mycrnotice%
\let\confname\myconfname%

\permission{Permission to make digital or hard copies of all or part of this work for personal or classroom use is granted without fee provided that copies are not made or distributed for profit or commercial advantage and that copies bear this notice and the full citation on the first page. Copyrights for components of this work owned by others than ACM must be honored. Abstracting with credit is permitted. To copy otherwise, or republish, to post on servers or to redistribute to lists, requires prior specific permission and/or a fee. Request permissions from permissions@acm.org. (c) 2014 Association for Computing Machinery. ACM acknowledges that this contribution was authored or co-authored by an employee, contractor or affiliate of the national government. As such, the Government retains a nonexclusive, royalty-free right to publish or reproduce this article, or to allow others to do so, for Government purposes only.}
\conferenceinfo{WiMobCity'14,}{August 11, 2014, Philadelphia, PA, USA.}
\copyrightetc{Copyright 2014 ACM \the\acmcopyr}
\crdata{978-1-4503-3036-7/14/08\ ...\$15.00.\\
http://dx.doi.org/10.1145/2633661.2633663}
\usepackage{graphicx}
\usepackage{amsmath}

\newtheorem{defn}{Definition}

\usepackage{hyphenat}

\usepackage{amsmath}
\usepackage{url}
\usepackage[numbers]{natbib}
\begin{document}
%
\title{Performance Comparison of Contention- and Schedule-based MAC Protocols in Urban Parking Sensor Networks} 

\author{
\and
Trista Lin, Herv{\'e} Rivano\\
       \affaddr{INRIA, Universit{\'e} de Lyon}\\
       \affaddr{INSA-Lyon, CITI-Inria}\\
       \affaddr{F-69621, Villeurbanne, France}\\
       \email{\{trista.lin, herve.rivano\}@inria.fr}
\and
Fr{\'e}d{\'e}ric Le Mou{\"e}l\\
       \begin{tabular}[t]{@{}c@{}}
       \affaddr{Universit{\'e} de Lyon}\\
       \affaddr{INSA-Lyon, CITI-Inria}\\
       \affaddr{F-69621 Villeurbanne, France}\\
       \end{tabular}\nobreak\qquad
       \begin{tabular}[t]{@{}c@{}} 
       \affaddr{Shanghai JianTong University} \\
       \affaddr{No. 800 DongChuan Rd.} \\
       \affaddr{Shanghai, China} \\
       \end{tabular} \\
       \email{frederic.le-mouel@insa-lyon.fr}
}



\maketitle

\begin{abstract}
Network traffic model is a critical problem for urban applications, mainly because of its diversity and node density. As wireless sensor network is highly concerned with the development of smart cities, careful consideration to traffic model helps choose appropriate protocols and adapt network parameters to reach best performances on energy-latency tradeoffs. In this paper, we compare the performance of two off-the-shelf medium access control protocols on two different kinds of traffic models, and then evaluate their application-end information delay and energy consumption while varying traffic parameters and network density. From the simulation results, we highlight some limits induced by network density and occurrence frequency of event-driven applications. When it comes to realtime urban services, a protocol selection shall be taken into account - even dynamically - with a special attention to energy-delay tradeoff. To this end, we provide several insights on parking sensor networks.
\end{abstract}



\newpage

\section{Introduction}


As the urban population is increasing, it brings the economic growth and the denser urban mobility\cite{UrbanMobility2012}. The first to be affected is the traffic congestion. Thanks to the smartphone technology, drivers can get diverse urban information simply from mobile apps. Thus, the availability and quality of urban information become the most important criteria for cities. User-generated urban information is the first proposed and enriches the information sources at different prospects. Various interesting information can also be shared according to users' sudden or periodic urban mobility. The published content could, however, be outdated or false because of insufficient participants or malicious users, as well as limited to human's observation. In view of this fact, wireless networked sensor devices help obtain more various types of information and assure of the accurate measurement. According to information types, sensor devices send updated information periodically or on-demand. That is, a network packet which consists of certain information, shall be treated with its corresponding priority in order to respect an acceptable information delay. Thanks to the increasing mobility need, some interesting urban services can be carried out by networked sensor devices. Among which, the traffic congestion is the greater thought at present and a huge percent of traffic jams are caused by the vehicles looking for parking spaces. So far as urban drivers are concerned, smart streetside parking system assisted by networked sensor devices is needed to shorten the parking search time and the parking distance from destinations. 

These networked parking sensor devices detect the availability of parking spaces and form a wireless sensor network (\textsc{Wsn}). Such a so-called parking sensor network (\textsc{Psn}) has the following characteristics: First, parking sensors are stationary, in-ground and scattered with a minimum adjacent distance. Second, the network topology is linear and limited by urban street layout. Third, the sensing area of each parking sensor does not have any intersection because of the lack of multiple detection. Fourth, packet generation rate depends on the vehicle's arrival and departure. Fifth, the availability parking information is the data of real-time parking service. Based on these, we see the importance of device lifetime considering their maintenance and latency while providing real-time service to urban citizens. Hence, we can say that \textsc{Psn} is a specialized form of \textsc{Wsn} and also inherits its energy and delay constraints.

To optimize network parameters for best performance, the network traffic is significant. Three mainstream traffic models are request, event and time-driven. Requ\-est-driven models are irrelevant for parking sensor networks since continuous recording is required for municipalities. Time-driven or periodic application is often used in testing the performance of network protocols because of its low dependence on the environment. Event-driven application is tricky due to its variety on different types of observed events. Considering different configurations of parking sensor networks, the \textsc{Mac} protocol, which deals with the bandwidth allocation, will be the first to be confronted. The \textsc{Wsn Mac} protocols are mainly classified as contention-based and schedule-based. Contention-based protocols provide a contention-based bandwidth allocation for sensor nodes and are widely discussed in urban applications. Schedule-based, also known as contention-free, protocols require at least one central node time-synchronized or asynchronized networks. Asynchronous method needs to perform a low power listening (\textsc{Lpl}) before transmitting each packet, thus its cumulative energy dissipation is not favorable for parking sensor network. From our survey, we are only interested in the impact of traffic intensity on time-synchronized \textsc{Mac} protocols. Our body of work is to simulate the traffic influence on stationary \textsc{Wsn} with the aim of improving the design of network architecture. Our contributions are summarized as follows: First, modeling of periodic and event-driven urban parking sensor applications by observing vehicle's arrival and departure. Second, energy and delay performance evaluation of two different applications through extensive experiments on urban scenarios, so as to compare two fundamental \textsc{Mac} protocols. Third, some key thresholds help to find a best configuration and highlight the importance to develop an adaptive \textsc{Mac} protocol which can distributedly detect the traffic intensity and switch between different configurations when required. Fourth, we also highlight some notbale issues while deploying multiple-hop parking \textsc{Psn}.

The remainder of this paper is structured as follows. In Section~\ref{sec:relatedworks}, we give a review on different MAC protocols. In Section~\ref{sec:networktrafficmodel} we introduce the network traffic model and the calculation of the information delay. In Section~\ref{sec:experience} we construct the urban environment and then perform the simulations by using contention-based and schedule-based \textsc{Mac} protocols. Finally, we summarize our works in Section~\ref{sec:conclusion}.


\section{Related works}
\label{sec:relatedworks}

Smart on-street parking application has received a lot of attention in recent years. Its main mission are to collect the realtime parking availability information and to disseminate the information to mobile users simply through a smart-parking app. Two types of collection methods are mobile and stationary. The former is to take advantage of vehicle's mobility to collect information along the route. In which, the most economical is crowdsourcing based mobile application. Therefore, it is obvious that crowdsourcing parking assistance system cannot provide a reliable realtime service required by municipalities\cite{6583499}. Neither does the mobile sensor side-mounted on a taxicab or bus for detecting an on-street parking map. For example, the ParkNet system in San Francisco\cite{Mathur:2010:PDS:1814433.1814448}, collects data with the location information from GPS receiver and then transmit it over a cellular uplink to the central server. Such a mobile parking sensor system requires much less installation, yet needs a longer average inter-polling time, to wit, 25 minutes for 80$\%$ of the cells in busier downtown area with only 300 cabs. Stationary collection method is to install on-site vehicle detection sensors\cite{6179195} to monitor the occupancy status of parking spaces. Based on this, large-scale road-side parking sensor network has been implemented in many cities, e.g., \textsc{SFpark} project\cite{SFparkreport} in San Francisco, \textsc{LA Express Park} in Los Angeles, \textsc{FastPrk} in Barcelona, \textsc{Connected Boulevard} in Nice and so forth. Among them, the efficiency of parking sensor network is the first concern. A lot of protocols have been proposed for urban \textsc{Wsn} applications. We sort them out as the following three groups.

\begin{itemize}
	\item \textsc{Contention-based} protocols are much widely studied in \textsc{Wsn} and generally based on or similar to \textsc{Csma}. When one node has a packet to send, it will have to struggle with the other competitors to get permitted of using the medium. The winner selection is somehow randomized. The synchronous \textsc{Mac} protocols are generally duty-cycled and require time-synchronized, such as \textsc{S-mac}, \textsc{T-mac}, \textsc{Conti}\cite{ctcontention2005}, \textsc{Sift}\cite{jamieson2006sift} and so forth. The state-of-the-art synchronization method is to be done through hardware or message exchange, and then a piggybacked acknowledge can be used to solve the clock shifting effect\cite{6488838}. The asynchronous versions use \textit{low-power listening}(\textsc{Lpl}) or its preamble-shortened approach to match up the transmission period between transmitter and receiver end, such as \textsc{B-mac}, \textsc{Wisemac}, \textsc{X-mac}, \textsc{Scp-mac}\cite{Ye:2006:UDC:1182807.1182839}, \textsc{Ri-mac}\cite{Sun:2008:RRA:1460412.1460414} and so on. Among them,\cite{4622710,6214027,thesisdequentin} compared the power dissipation between asynchronous and synchronous contention-based protocol. In which, \textsc{Lpl} method is interesting in very low traffic intensity (less than one packet per day) or dynamically changing topologies. Otherwise, synchronous protocol outperforms asynchron\-ous one. The drawback of such a protocol is the packet collision causing by increasing network density and hidden terminal. 
	
	\item \textsc{Schedule-based} protocols are generally centralized and suitable for static topologies. Assigned nodes play the master role to allocate slotted network resource to their slaves. The mechanism is generally based on \textsc{Tdma} or \textsc{Cdma}. The clock of each node must be time-synchronized. Scheduled slots can be fixed or on-demand. Some noted protocols are like \textsc{Drand}\cite{Rhee:2006:DDR:1132905.1132927}, \textsc{Leach}\cite{926982}, \textsc{Trama}\cite{Rajendran:2003:ECM:958491.958513} and \textsc{Tsmp}\cite{Pister08tsmp:time}. They use \textsc{Tdma} as the baseline \textsc{Mac} scheme, and then take \textsc{Csma}, \textsc{Aloha} or \textsc{Cdma} for improving these join/leave/synch-ronize messages. The drawbacks are firstly not easy to adapt to the dynamics of network, and secondly the slower response of the centralized control while adapting the schedule to the traffic variation. 
		
	\item \textsc{Hybrid contention- \& schedule-based} is to combine the advantages of both protocols in order to reach the best performance. \textsc{Z-mac}\cite{4453818} behaves like \textsc{Csma} under low competition and under high competition, like \textsc{Tdma}. \textsc{Funneling-mac}\cite{Ahn:2006:FLS:1182807.1182837} uses \textsc{Csma} as the baseline, and changes to \textsc{Tdma} while receiving on-demand beaconing from the sink, that is to say, nodes close to the sink performs \textsc{Tdma}. \textsc{Funneling-mac} works in the application of data collection. \textsc{iQueue-mac}\cite{6644967} runs in \textsc{Csma} in light load and then uses queue-length piggybacking as accurate load information to ask for additional variable \textsc{Tdma} time slots if needed. 	 
\end{itemize}

The occupancy and vacant durations are different to get due to the constraint of real \textsc{Psn} implementation. The arrival and departure were generally assumed Poisson distributed, and the occupancy rate of the parking system can be analyzed by Markov process. However, the newest report in Santander shows that the vehicle's occupied time can be described by Weibull distribution\cite{SmartSantanderReport}. Therefore, which type of MAC protocol suites best the urban smart parking application in terms of network density and traffic intensity?  In this paper, we configure several different scenarios with different network density, traffic intensity and two mainstream \textsc{Mac} protocols in order to find the best configuration while constructing urban WSN applications.

\section{Network traffic model}
\label{sec:networktrafficmodel}

In urban sensor network, sensors are often stationary to monitor certain events, and thence the packet transmission shall always happen and never end. In other words, the key point of modeling an event-driven application will be to find an appropriate distribution to define the traffic interval, viz inter-event interval. Next, we present the modeling of the event occurrence and the definition of information delay with respect to urban smart parking applications.

\begin{figure}[!t]
  \centering
  \includegraphics[width=\linewidth]{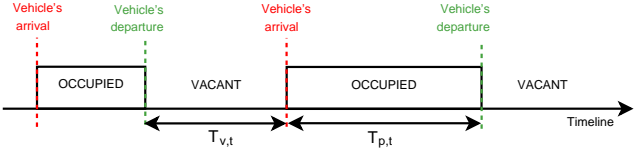}
  \caption{\footnotesize{Timeline of sensor's occupancy status}}
  \label{fig:occupancy-status}       
\end{figure}

\subsection{Vehicle's arrival and departure}

In parking sensor networks, the main observed events are vehicles' arrivals and departures. For any parking sensor, each vehicle arrival accompanies exact by one departure prior to next arrival. To model it, we first look at the event occurrence sequence on one parking sensor. We suppose each parking sensor is precise enough and provides merely two status, namely occupied and vacant. The occupied interval from vehicle's arrival to departure is so-called parking time $T_{p,t}$, conversely, the vacant interval is available time $T_{v,t}$. During which, each sensor detects vehicle's presence or absence at a given time $t$, shown in Figure~\ref{fig:occupancy-status}. Both $T_{p,t}$ and $T_{v,t}$ proper shall be described by a fitting distribution in order to approximate their randomness. From Vlahogianni's report\cite{SmartSantanderReport}, the massive real-time parking availability data, obtained by an on-street parking sensor network in Santander, shows that the occupied duration is best described by a Weibull distribution. By assuming that $T_{p,t}$ and $T_{v,t}$ are both Weibull distributed with rate parameters $\lambda$ and $\mu$ and shape parameters $\alpha$ and $\beta$, we have: 

\begin{itemize}
  \item $Pr(T_{p,t}=X) = \frac{\alpha}{\lambda^{\alpha}}X^{\alpha-1}\mathrm{e}^{-(\frac{X}{\lambda})^{\alpha}}$ stands for the probability of choosing a occupancy time $X$. 

  \item $Pr(T_{v,t}=Y) = \frac{\beta}{\mu^{\beta}}Y^{\beta-1}\mathrm{e}^{-(\frac{Y}{\mu})^{\beta}}$ stands for the probability of choosing a vacant time $Y$. 
\end{itemize}

\begin{figure}[!t]
\begin{minipage}{.49\linewidth}
 \centering
 \includegraphics[height=\linewidth, angle=-90]{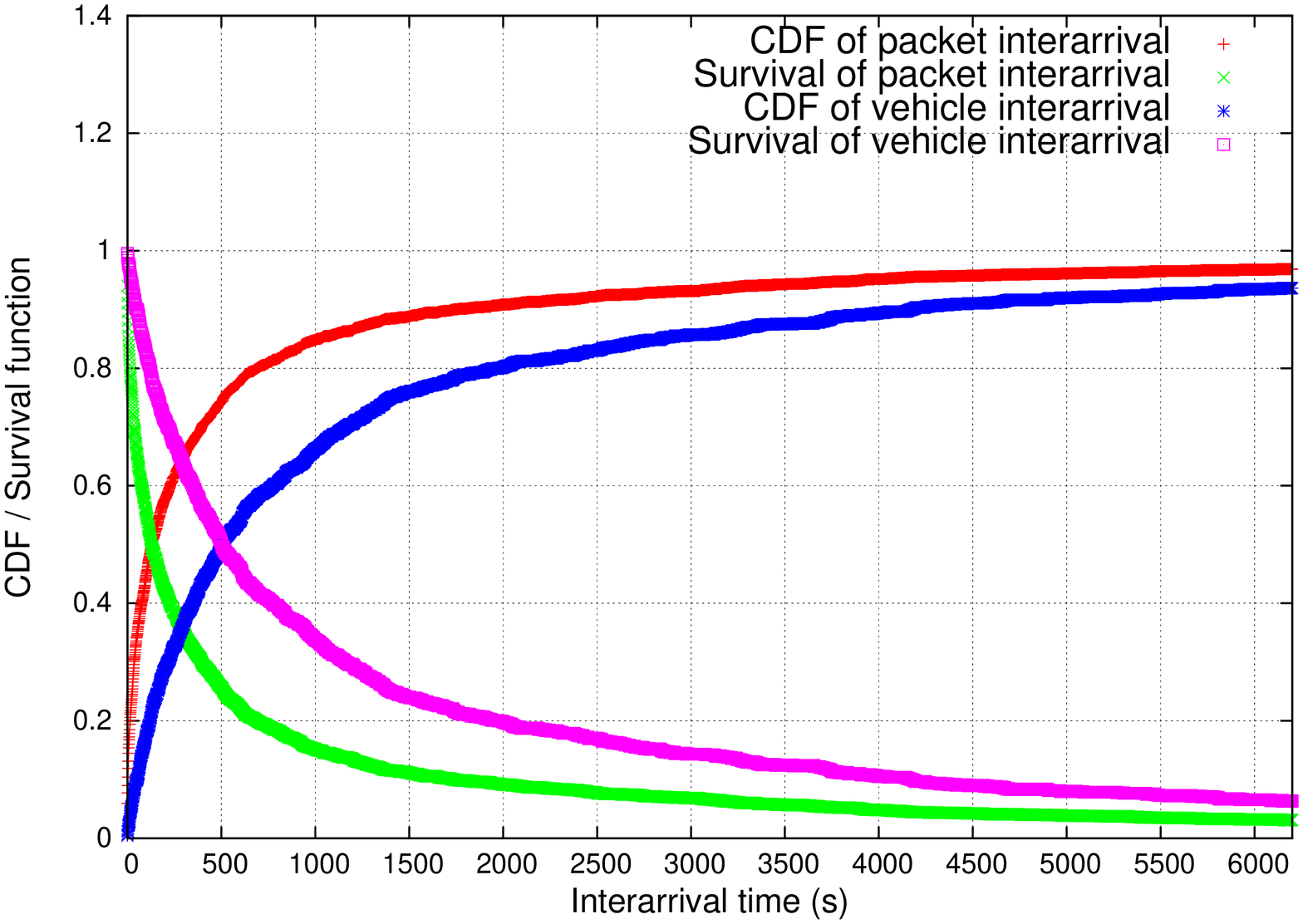}
 \caption{\footnotesize{The CDF and survival function of vehicle and packet interarrival times of one in-ground parking sensor}}
 \label{fig:cdf-one-node} 
\end{minipage}
\hfill
\begin{minipage}{.49\linewidth}
 \centering
 \includegraphics[height=\linewidth, angle=-90]{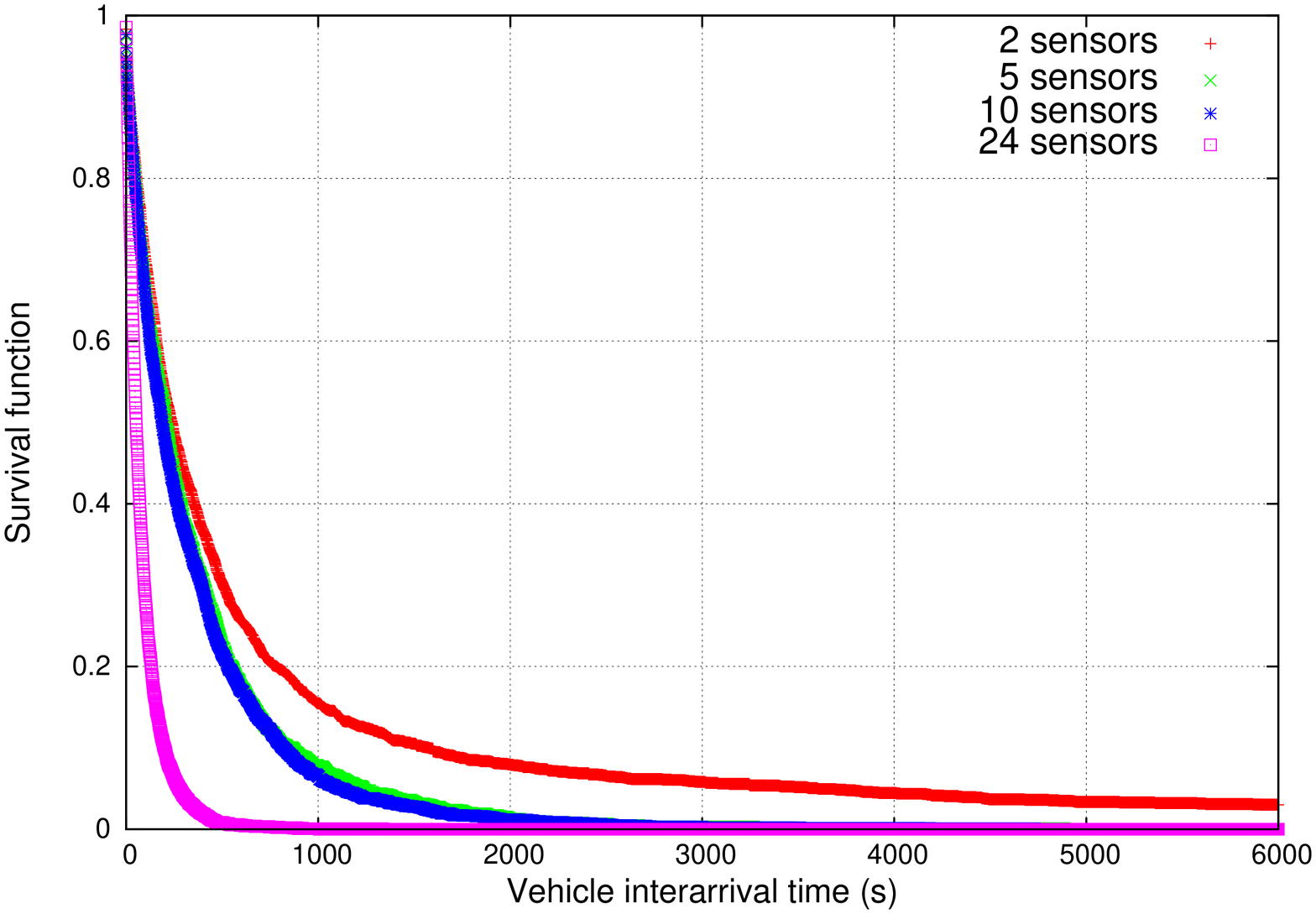}
 \caption{\footnotesize{The survival function of vehicle interarrival times of 2, 5, 10 and 24 in-ground parking sensors}}
 \label{fig:cdf-n-nodes}             
\end{minipage}
\end{figure}

The vehicle's interarrival duration can be expressed as $Z=T_{p,t}+T_{v,t}$ in which, one turnover is counted. The occupancy rate will be $E[T_{p,t}/(T_{p,t}+T_{v,t})]= \lambda \Gamma(1+\frac{1}{\alpha})/(\lambda \Gamma(1+\frac{1}{\alpha})+\mu \Gamma(1+\frac{1}{\beta}))$. From literature, Pareto and Weibull distribution have been discussed to describe the burstiness of network traffic. While looking at the Figure 5 in \cite{SmartSantanderReport}, the occupancy duration in four different regions follow the Weibull distribution with the shape parametes approximately 0.4$\sim$0.7. When the shape parameter is smaller than 1, the occupancy duration is heavy-tailed\cite{6662948}. We took the Weibull parameters in \cite{SmartSantanderReport} and then get its CDF and survival of $Z$ in Figure~\ref{fig:cdf-one-node}. Hence, if $T_{p,t}$ and $T_{v,t}$ are both heavy-tailed, so is $T_{p,t}+T_{v,t}$. From \cite{Mitov2006555}, the cdf $F(t)$ of interarrival time $Z$ has regularly varying tail such that: $1-F(t) \sim t^{-\epsilon}L(t)$ as $t\rightarrow \infty$. $L(t)$ is a slowly varying function at infinity, that is, $\lim\nolimits_{t\rightarrow \infty}L(tx)/L(t)=1 \; \forall x>0$. From Mitov's model, the distribution function of the interarrival times is Weibull with $0 < \epsilon \leq 1$, the shape parameter $\nu = 1 - \epsilon$ and scale parameter $\gamma = C\sin\frac{\pi\epsilon}{\pi(1-\epsilon)}$, provided that $Nt^{1-\epsilon}\rightarrow C$, $0<C<\infty$ and $N$ is the number of traffic sources, namely parking sensors. If we increase the amount of observed parking sensors, i.e., $N$, the vehicle interarrival time is shown in Figure~\ref{fig:cdf-n-nodes} and the shape parameter is approximating to 1. The count model based on Weibull interarrival times is studied in \cite{EricBradlow}. The probability of $k$ arriving vehicles in a given interval is calculated as below:

\begin{equation}
P(N(t)=k)=\sum_{j=k}^{\infty}\frac{(-1)^{j+k}(\frac{t}{\gamma})^{\nu j} \Delta_{j}^{k}}{\Gamma(\nu j+1)} \;\; k = 0,1,\cdots
\label{eq:prob_of_k_vehicle_arriving}
\end{equation}

where $\Delta_{j}^{k+1} = \sum\nolimits_{m=k}^{j-1} \Delta_{m}^{k} \frac{\Gamma(\nu j - \nu m +1)}{\Gamma(j - m +1)}$ for $k=0,1, 2,\cdots$ and $j=k+1,k+2,k+3,\cdots$. $\Delta_{j}^{0}=\frac{\Gamma(\nu j+1)}{\Gamma(j+1)} \; j=0,1,2,\cdots$. In a business area or on weekdays, the average occupied time can be shorter according to the area hourely activities or parking policy. Its impact to network traffic will have to be considered as well.

\subsection{Network traffic models}

\subsubsection{Event-driven traffic}

\begin{figure}[!t]
\begin{minipage}{.52\linewidth}
 \centering
 \includegraphics[width=\linewidth]{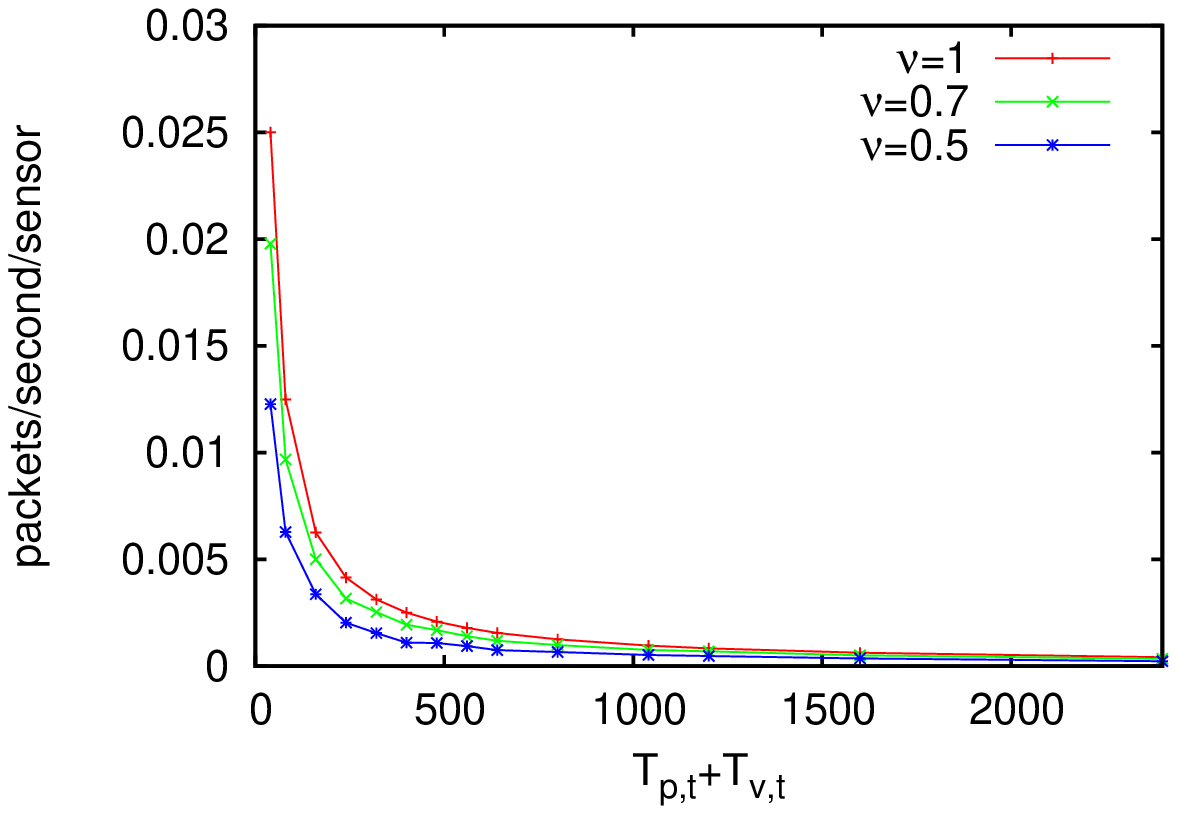}
 \caption{\footnotesize{Generated packets by event-driven traffic model while all sensors have the same scale parameters($\gamma$).}}
 \label{fig:trigger-app} 
\end{minipage}
\hfill
\begin{minipage}{.46\linewidth}
 \centering
 \includegraphics[width=\linewidth]{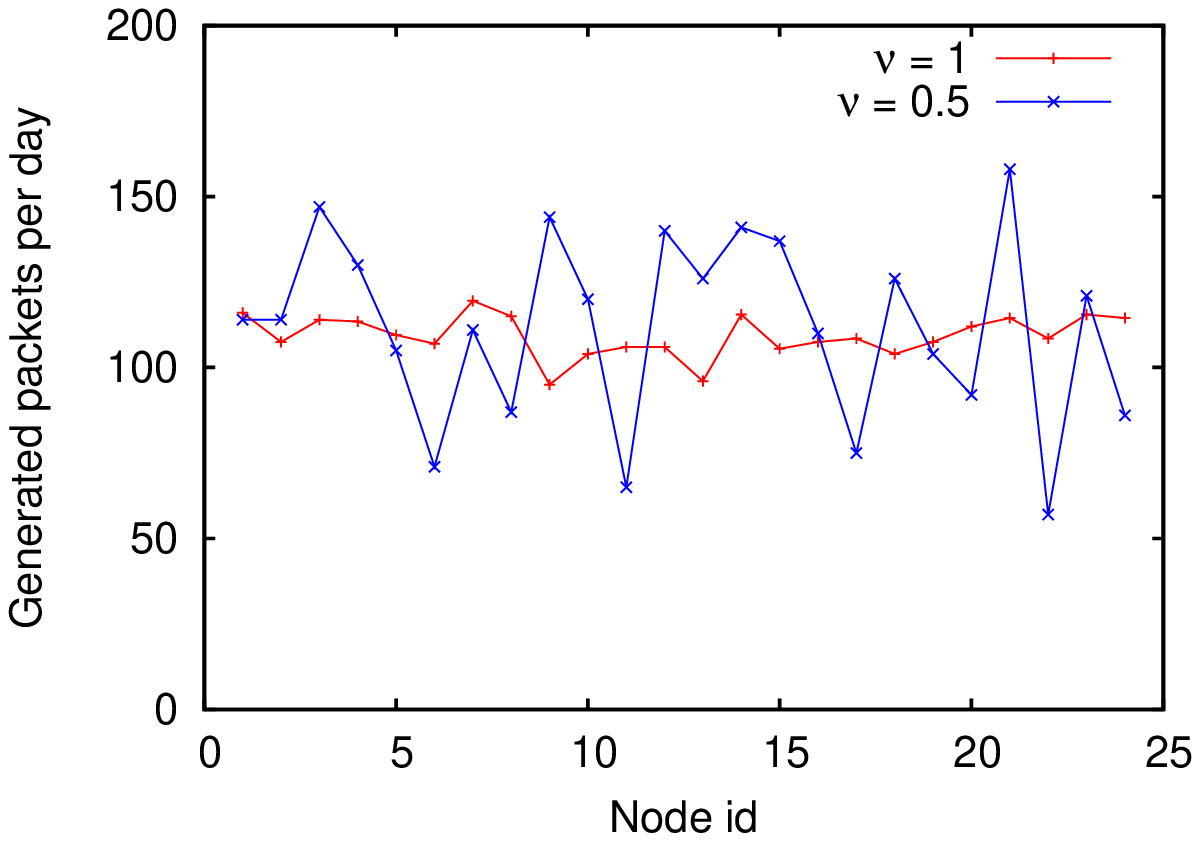}
 \caption{\footnotesize{Traffic difference between sensor nodes while all the sensors have the same mean of interarrival time($\frac{Z}{2}$) but different shape parameters($0.5$ and $1$).}}
 \label{fig:trafficdifference}             
\end{minipage}
\end{figure}

\begin{figure}[!t]
\begin{minipage}{\linewidth}
 \centering
 \includegraphics[height=\linewidth, angle=-90]{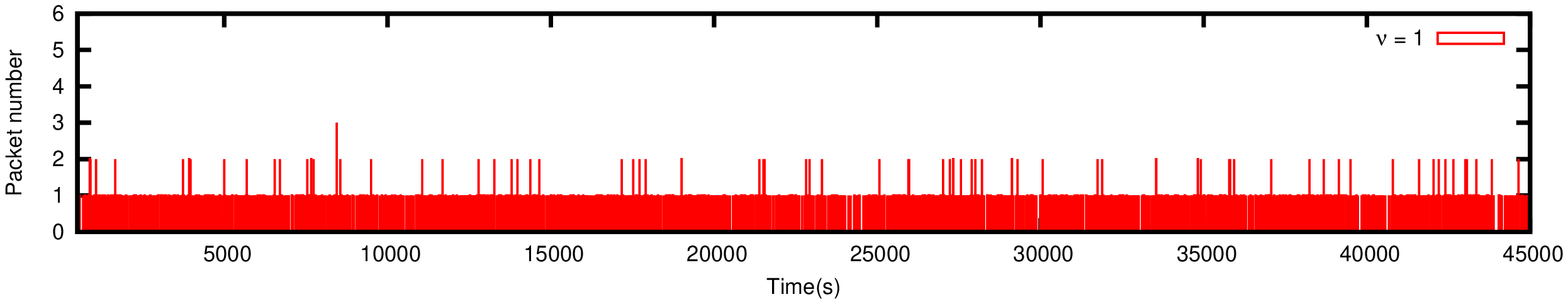}
\end{minipage}
\hfill
\begin{minipage}{\linewidth}
  \centering
  \includegraphics[height=\linewidth, angle=-90]{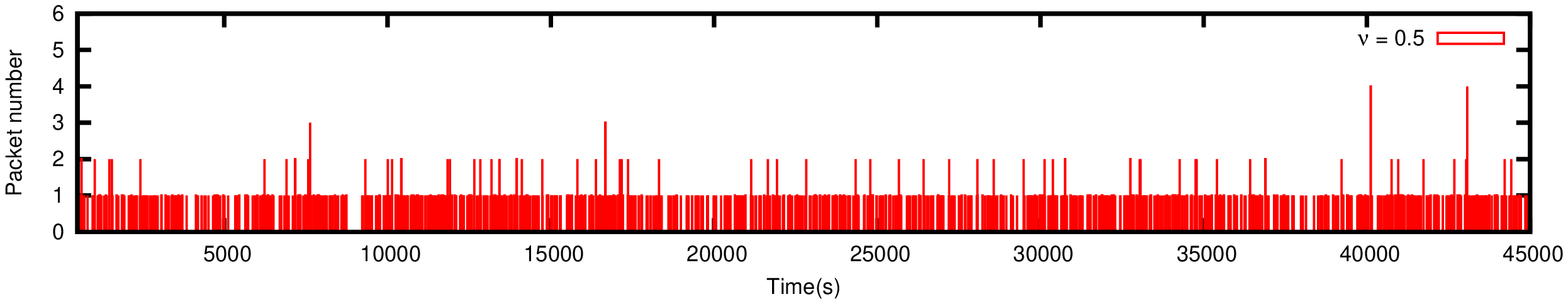}
\end{minipage}
\caption{\footnotesize{Timeline of Network Traffic when having the same scale parameter but different shape parameters $0.5$ and $1$}}
 \label{fig:burst-1-0.5}   
\end{figure}

The event occurrence frequency has a great impact on event-driven applications. Each parking sensor sends one packet while one vehicle arrives and another when it leaves, that is, sensor sends two packets every $T_{p,t}+T_{v,t}$ which is the interarrival time of vehicles. The average interarrival of network packets from parking sensor $i$ is $\frac{1}{2}(\lambda_{i} \Gamma(1+\frac{1}{\alpha_{i}})+\mu_{i} \Gamma(1+\frac{1}{\beta_{i}}))$ per second. While considering $N$ parking spaces, the count model of total generated network packets is:

\begin{equation}
C_{N}(\Delta t) = \sum\limits_{i=1}^{N} \frac{2 \Delta t}{\lambda_{i} \Gamma(1+\frac{1}{\alpha_{i}})+\mu_{i}\Gamma(1+\frac{1}{\beta_{i}})}
\label{eq:count-packet-model}
\end{equation}

Where $\Delta t$ is the observation period. When $C_{N_{1}}(\Delta t)= C_{N_{2}}(\Delta t)$, the packet and vehicle interarrival durations are the same. However, in such a case, even though $N_1=N_2$, we can still find several sets of different combinations of $\lambda_{i}$, $\alpha_{i}$, $\mu_{i}$ and $\beta_{i}$. These parameters decide the variation of $T_{p,t}$ and $T_{v,t}$ which give the information about the utilisation of street parking.

Exponential distribution has been widely used for simulating the vehicle's arrival. It is one shape of Weibull distribution while shape parameter is equal to $1$. Hence, we try to compare the differences while applying Weibull distribution in the network traffic model with respect to Exponential distribution. Figure~\ref{fig:trigger-app} shows that the packet generation rate is strongly decided by the sum of $T_{p,t}$ and $T_{v,t}$ from equation~(\ref{eq:count-packet-model}), and also proportional to the shape parameter($\nu$) when the scale parameter($\gamma$) is fixed. The reason of reducing packet generation rate is caused by part of longer parking occupancy time, which is well described by heavy-tailed distribution. 

Then we fix the average of vehicle interarrival time and compare two different shape parameters $0.5$ and $1$. In Figure~\ref{fig:trafficdifference}, we see that the non-uniform property is more obvious when the shape parameter is smaller than 1. This is because the infinite mean and infinite variance which make a larger variations among parking sensors. Moreover, what is the impact to the network coordinator? Figure~\ref{fig:burst-1-0.5} shows the network load while managing $24$ parking sensors. We can see clearly that when the shape parameter is $0.5$, the burstiness of network traffic is more significant but might not be considered in Exponential distribution.

\subsubsection{Periodic/Time-driven traffic}

On the contrary, periodic application is only affected by the traffic interval $\omega$ instead of event occurrence frequency. The amount of generated packets is inversely proportional to the traffic interval. While using periodic traffic model, the amount of network traffic is unaffected by the sensory information. It simply sends out a packet with the current time-stamped status when the time is up.

\begin{defn}
Information delay $T_{delay}$ is the required time for knowing a changed occupancy status of a parking sensor.
\end{defn}


In event-driven application, each sensor sends out the updated information at once when detecting any event, namely, application delay is almost zero. Periodic application is subject to the traffic interval so that an application delay shall be added up.

\section{Parking sensor network experiments}
\label{sec:experience}

\subsection{Bandwidth allocation methods}
\label{subsec:ba-method}

Before starting the simulation, we compared several different MAC protocols. From the literature, we find that to choose a contention- or schedule-based \textsc{Mac} protocol has been the subject of much controversy. To compare their efficiency, we evaluate two off-the-shelf \textsc{Mac} protocols: duty-cycled \textsc{Tdma} for schedule-based and duty-cycled \textsc{Csma/ca} for contention-based. From the reason mentioned in section~\ref{sec:relatedworks}, both of them are running under a time-synchronized environment. To minimize the idle listening period in each data slot, each transmitter sends a very small beacon to reserve an appointment before starting the data transmission. If the reservation fails, the transmitter will put the packet into the queue, turn off its radio and wait for the next time slot. Instead of wasting energy to do a vain transmission, sensor nodes prefer to evaluate the receiver's availability through these very small beacons.

\subsubsection{Schedule-based bandwidth allocation}

Considering the sensor network is often bandwidth-limited, the only medium resource is time divisions in a single-channel scenario. Each node arrived in the network will send a request in order to be preassigned to one partition of medium resource to transmit their packets. While the network coordinator does not know in advance the traffic model and node position, it preallocates an equal partition to all the nodes in his subnet. If a node has no packets to send in its term, the others still cannot seize this occasion to send their packets out. 

\begin{figure}[ht]
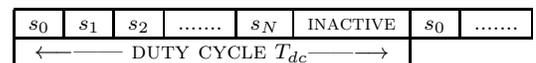

\centering
\begin{tabular}{|c|c|c|c|c|c|c|c|}
\hline
\small{$s_{0}$}  & \small{$s_{1}$} & \small{$s_{2}$} & \small{.......} & \small{$s_{N}$} & \textsc{\small{inactive}} & \small{$s_{0}$}  & \small{.......} \\
\hline
\multicolumn{6}{|c|}{$\longleftarrow$------ \textsc{duty cycle} \small{$T_{dc}$}------$\longrightarrow$}
\end{tabular}
\caption{\footnotesize{Duty cycle of schedule-based B.A.}}
\label{fig:duty-cycle-tdma}
\end{figure}

The duty cycle comprises an inactive period and at least $N+1$ time slots for $N$ parking sensors and $1$ coordinator, shown in Figure~\ref{fig:duty-cycle-tdma}. Each sensor can only send its packet on its pre-assigned time slots. The duration of duty cycle can be calculated by $T_{dc} = T_{slot}*(N+1) + T_{inactive}$. If the packet transmission fails on the current time slot, the next retrial will be in $T_{slot}(N+1)$ seconds. The advantages of schedule-based B.A. are the nearly very little packet collision rate and lower energy consumption since nodes only send beacons in certain slots. If sensor nodes' traffic model is given, schedule-based B.A. can optimize the resource assignment to reach a better performance. The drawbacks are that many time slots are wasted and a longer delay time is caused, also the urban traffic model is dynamic and time-variant. Else, if there is a new node which intends to join this network, the gateway will have to reallocate the resource while there is no enough time slots.

\subsubsection{Contention-based bandwidth allocation}

Contention-based B.A. uses competition based medium access control due to the inflexible resource allocation previously mentioned. The principle is that the node gets its partition of network resource when it asks for. If more than two nodes declare their demands, a competition will be held to choose who is the current transmitter. The transmitter candidates randomly choose a waiting time for sending the reservation beacon and then listen to the medium before the waiting time expires. If these candidates hear any beacon in the meantime, they will turn off the radio and wait for next resource allocation. Otherwise, the candidate will send out a beacon message to announce its transmission. 

\begin{figure}[ht]
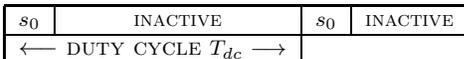

\centering
\begin{tabular}{|c|c|c|c|c|c|}
\hline
\multicolumn{2}{|c|}{\small{$s_{0}$}} & \textsc{\small{inactive}} & \multicolumn{2}{|c|}{\small{$s_{0}$}} & \textsc{\small{inactive}} \\
\hline
\multicolumn{3}{|c|}{$\longleftarrow$ \textsc{duty cycle} \small{$T_{dc}$} $\longrightarrow$} \\
\end{tabular}
\caption{\footnotesize{Duty cycle of contention-based B.A.}}
\label{fig:duty-cycle-802.15.4}
\end{figure}

The duty cycle contains the competition, data transmission and inactive periods, shown in Figure~\ref{fig:duty-cycle-802.15.4}, $T_{dc} = T_{slot} + T_{inactive}$. The advantages of contention-based resource allocation are the better use of network resource and a short network delay. The drawbacks is the inevitable packet collision which causes arbitrarily high energy consumption and latency on grounds of endless competitions triggered by high node density. If there is a new node which intends to join this network, it will just join the competition and increase the packet collision rate.

\subsection{Simulation experiences}

\begin{figure}[!t]
\centering
\includegraphics[width=\linewidth]{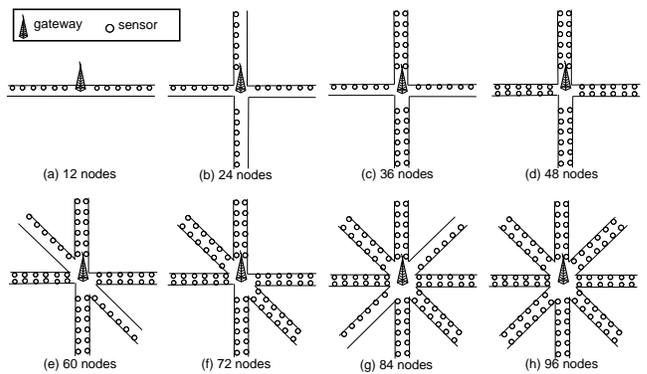}
\caption{\footnotesize{Topologies of different numbers of sensors. Increase the node density by adding one more line or one more side of curb parking.}}
\label{fig:topology}
\end{figure}

From the real implementation in SFpark project\cite{SFparkreport}, we find that the network coordinators (router or gateway) are always installed in crossroads considering the maximum coverage and the facility of relays. Hence, we first take one small cell in a crossroad with different network density. The topologies are depicted in Figure~\ref{fig:topology}. Our simulations, performed with the \textsc{WSNet} simulator\cite{WSNet}, use the simulation parameters in table~\ref{tab:SimulationParameters}. We set the slot durations are both $0.1$ seconds in order to have the same active time and so does the throughput.

\begin{table}[!t]
\vspace{-3mm}
\caption{Simulation parameters}
\label{tab:SimulationParameters}
\small
\centering
	\begin{tabular}{|l|l|l|l|}
	\hline
	\multicolumn{2}{|l|}{Simulation time: 86400 seconds} & \multicolumn{2}{|l|}{Batch: \#20}\\	
	\hline
  \multicolumn{2}{|l|}{Sensor number: 12 $\sim$ 96} & \multicolumn{2}{|l|}{Gateway number: 1}  \\
	\hline
	\multicolumn{4}{|l|}{Distance between two adjacent sensors: 5 meters} \\
	\hline
	\multicolumn{4}{|l|}{Distance from the gateway to its vicinal sensors: 10 meters}\\

	\hline
	\multicolumn{2}{|l|}{Transmit power output 3 dBm} & \multicolumn{2}{|l|}{Receive sensitivity -85 dBm} \\
	\hline
	\multicolumn{2}{|l|}{Data rate 250 kbps} & \multicolumn{2}{|l|}{802.15.4 Radio} \\
	\hline
	$P_{tx}$ 65.7 mW  & $P_{rx}$ 56.5 mW & $P_{cs}$ 55.8 mW& $P_{off}$ 30 $\mu$W \\
  \hline
  \multicolumn{2}{|l|}{$E_{radio.switch}$ 0.16425mJ} & \multicolumn{2}{|l|}{Packet size 84 bytes} \\
	\hline      
  \multicolumn{4}{|l|}{MAC: duty-cycled schedule \& contention-based bandwidth} \\
  \multicolumn{4}{|r|}{ allocation, slot duration=0.1s. Retransmission\&piggyback.} \\
  \hline  
  \multicolumn{4}{|l|}{Application: Event-driven - $\lambda$, $\mu$ \& $\alpha=\beta=0.5$ , Periodic - $\omega$} \\
	\hline
	\multicolumn{4}{|l|}{Routing: gradient\cite{Watteyne-5425543}} \\
	\hline
	\multicolumn{4}{|l|}{Propagation: Corner pathloss($\lambda = 0.125$)\cite{1492678}+Rayleigh fading} \\
  \hline
	\end{tabular}
	\vspace{-4mm}
\end{table}

\subsubsection{Impact of network density}

\begin{figure}[!t]
\begin{minipage}{.49\linewidth}
  \centering
  \includegraphics[width=\linewidth]{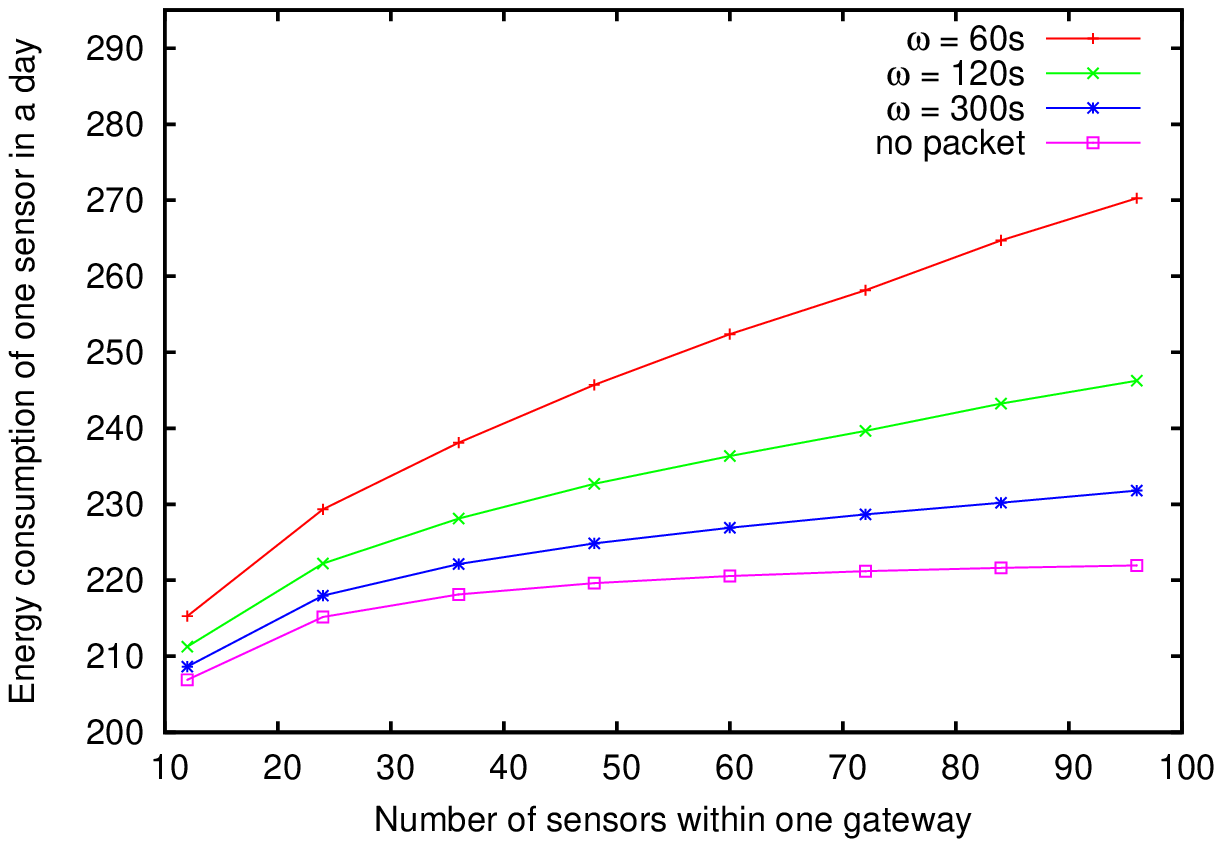}
\end{minipage}
\hfill
\begin{minipage}{.49\linewidth}
 \centering
 \includegraphics[width=\linewidth]{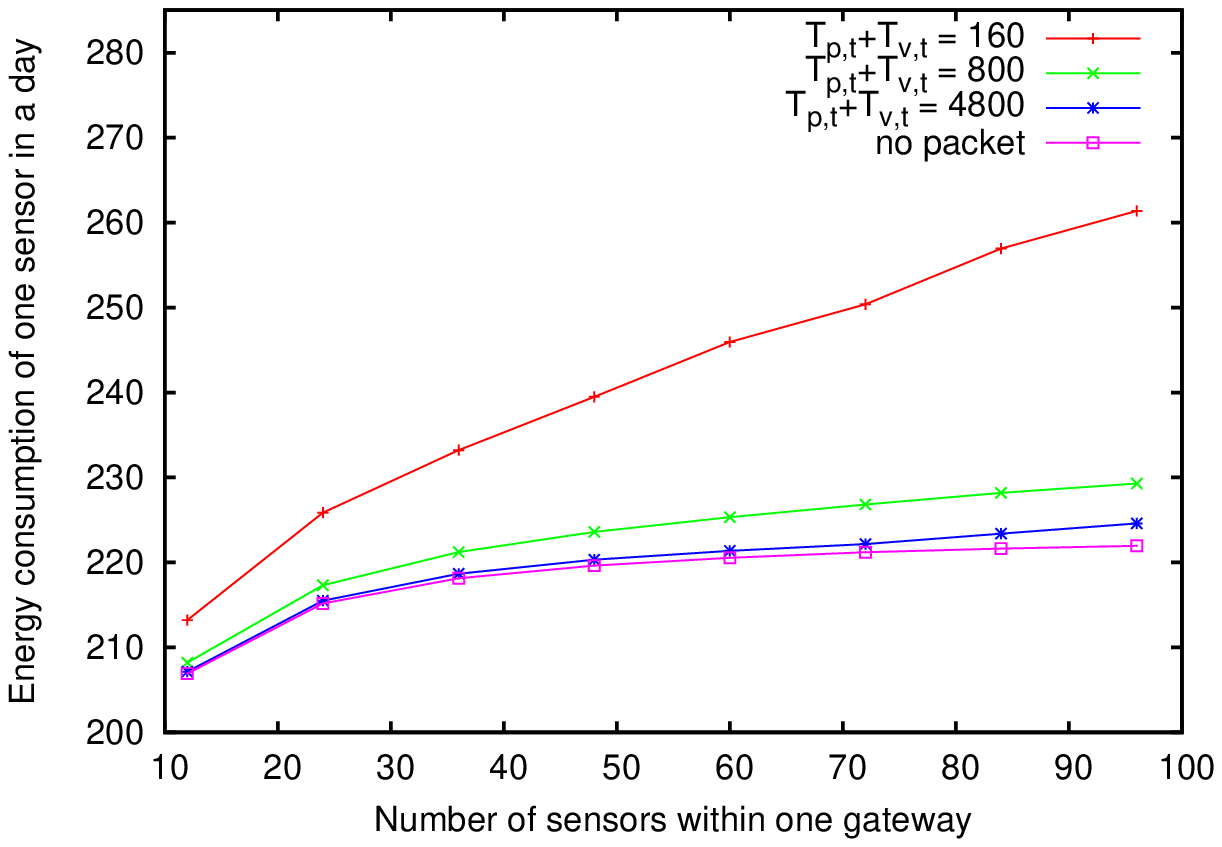}
\end{minipage}
\caption{\footnotesize{Per-node energy consumption of (left) periodic and (right) event-driven applications while varying $N$ and traffic parameters under \emph{schedule}-based B.A.}}
\label{fig:tdma-time-event-energy-per-node}
\begin{minipage}{.49\linewidth}
  \centering
  \includegraphics[width=\linewidth]{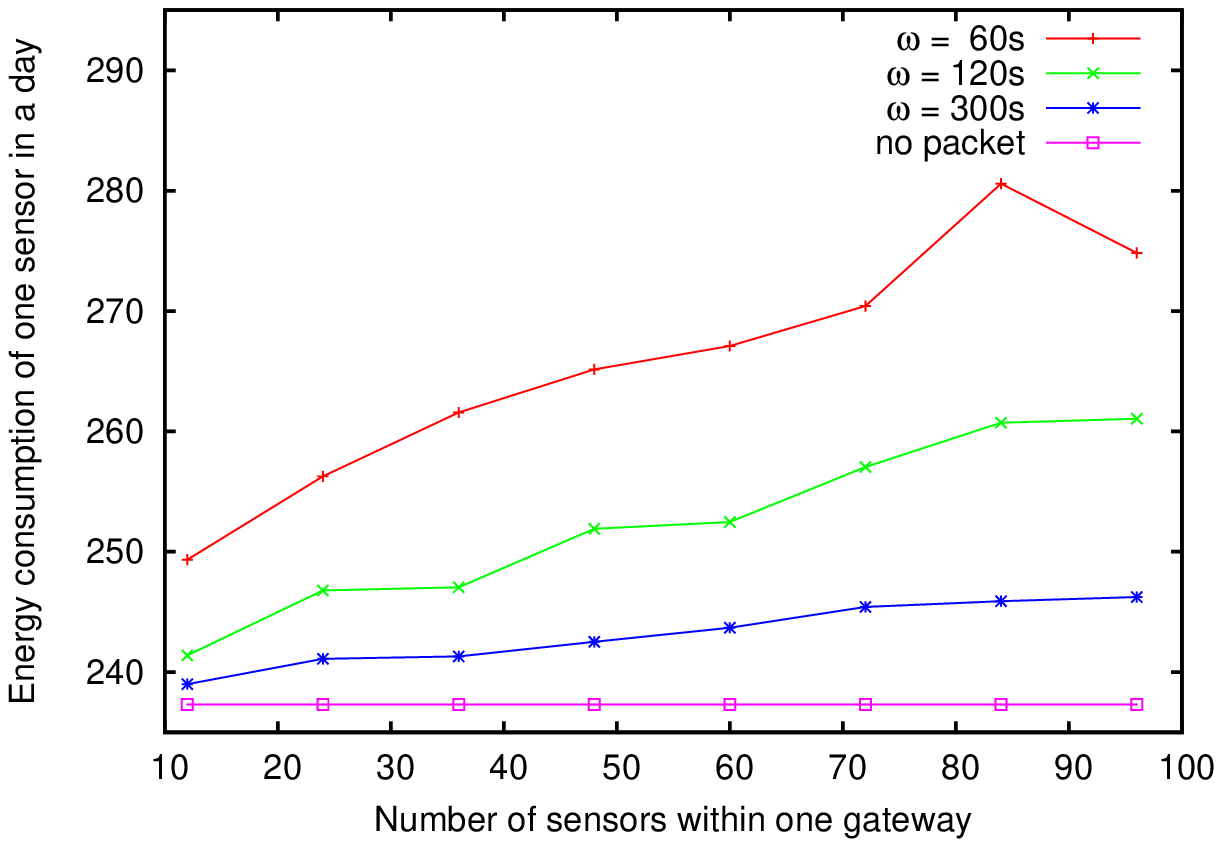}
\end{minipage}
\hfill
\begin{minipage}{.49\linewidth}
 \centering
 \includegraphics[width=\linewidth]{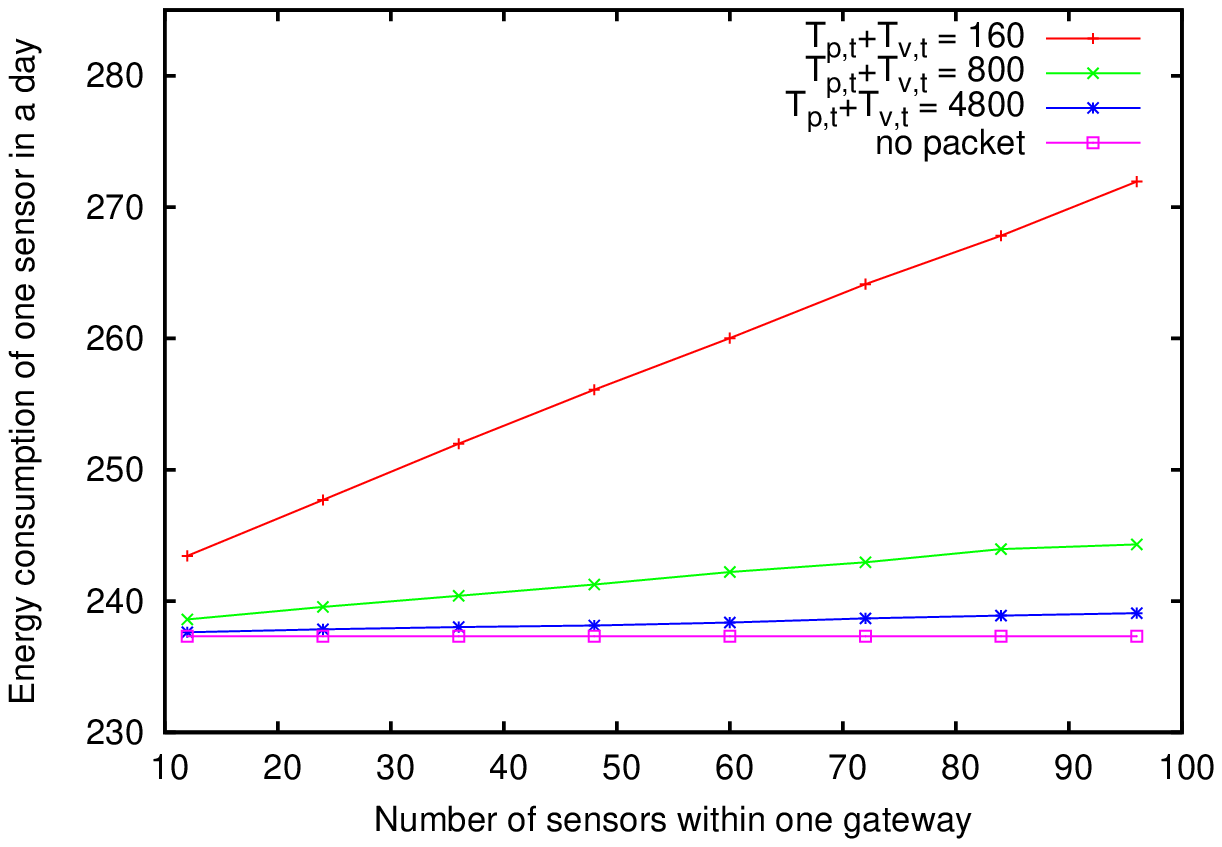}
\end{minipage}
\caption{\footnotesize{Per-node energy consumption of (left) periodic and (right) event-driven applications while varying $N$ and traffic parameters under \emph{contention}-based B.A.}}
\label{fig:802.15.4-time-event-energy-per-node}
\begin{minipage}{0.6\linewidth}
 \centering
 \includegraphics[width=\linewidth, angle=-90]{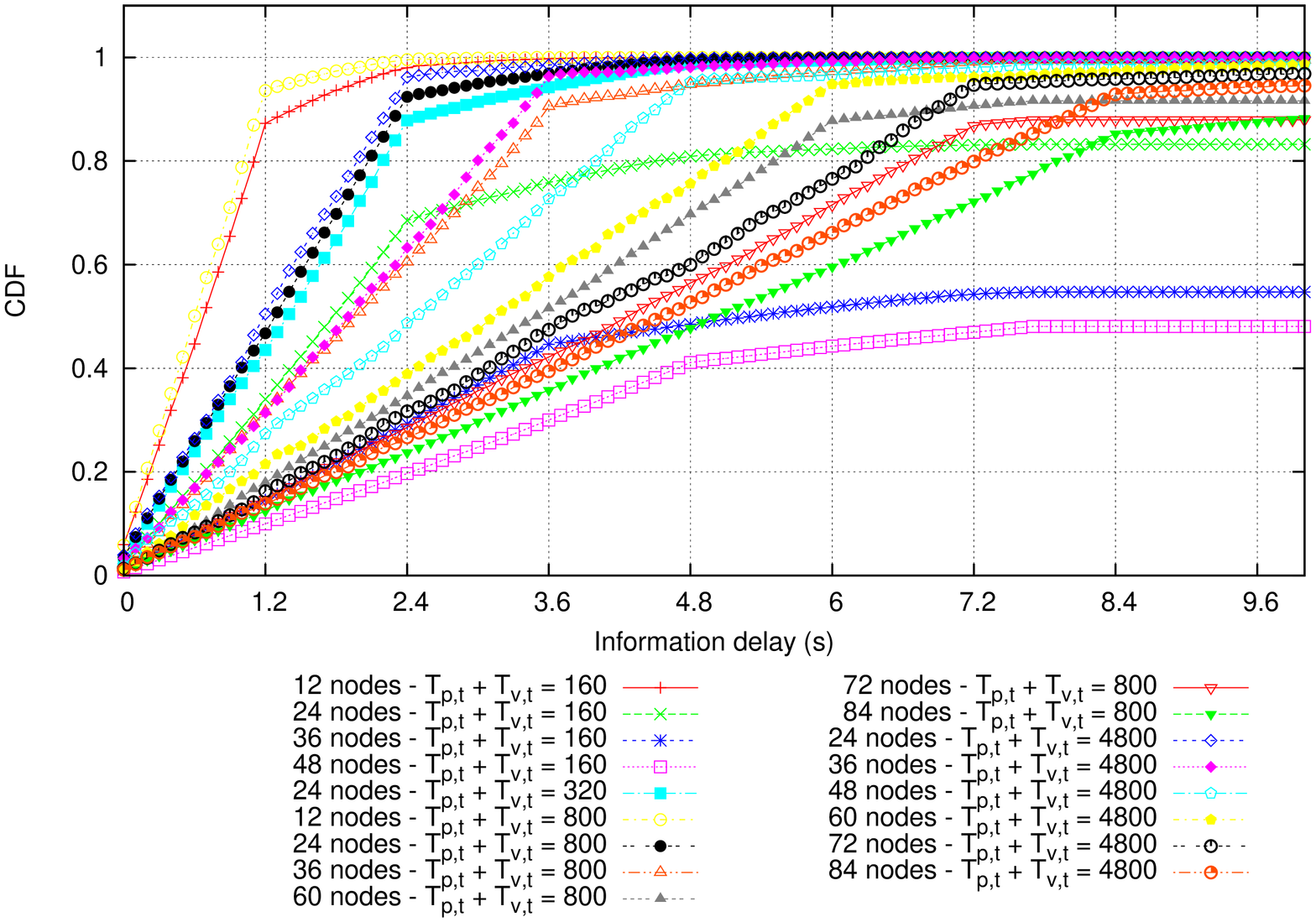}
\end{minipage}
\caption{\footnotesize{Information delay of \emph{event}-driven application while varying traffic parameters and node number under \emph{schedule}-based B.A.}}
\label{fig:tdma-event-delay-different-nodes}   
\begin{minipage}{0.6\linewidth}
 \centering
 \includegraphics[width=\linewidth, angle=-90]{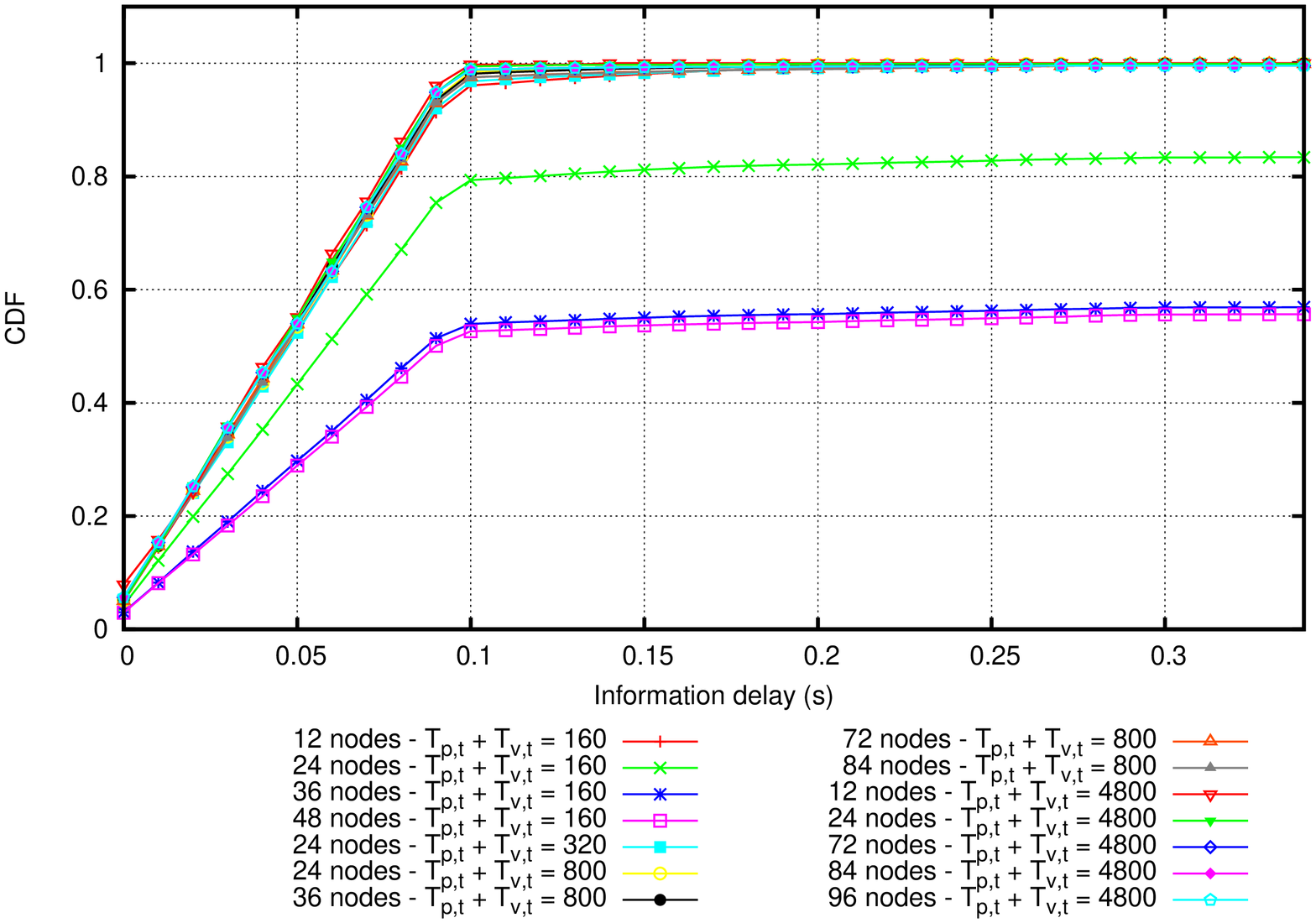}
\end{minipage}
\caption{\footnotesize{Information delay of \emph{event}-driven application while varying traffic parameters and node number under \emph{contention}-based B.A.}}
\label{fig:802.15.4-event-delay-different-nodes}
\end{figure}

Figures~\ref{fig:tdma-time-event-energy-per-node} and~\ref{fig:802.15.4-time-event-energy-per-node} shows the per-node energy consumption using periodic and event-driven network traffic models. The x-axis stands for the sensor number in the network. In schedule-based B.A., the minimum energy depletion is required to arrange the schedule. As the network dimension increases, each sensor also consumes more energy. Contention-based B.A. suffers from packet collision when the network competitors increase. Else, periodic traffic, which generates one packet per $60$ seconds, has too much interference and cannot send more packets after there are more than $84$ network competitors. Figures~\ref{fig:tdma-event-delay-different-nodes} and~\ref{fig:802.15.4-event-delay-different-nodes} show the information delay of event-driven application under schedule- and contention-based B.A.. The information delay of periodic application is strongly related to the traffic interval and not appropriate for real-time urban application. We will show the result in next subsection. Event-driven application is more sensitive to the bandwidth allocation methods and generally proportional to the duration of duty cycle $T_{dc}$. From the result, we know that: First, the probability of a packet arriving in $k^{th}$ duty cycle is $ p_{k}\prod\nolimits_{i=1}^{k-1} (1-p_{i})$ where $p_{i}$ is the probability that the packet is received in $i^{th}$ cycle. Second, when a packet arrives in $k^{th}$ duty cycle, the arriving time point is uniformly distributed with the mean $\frac{T_{dc}}{2}$. The expectation of information delay of event-driven application is calculated as below:


\begin{align}
\label{eq:infodelay-event}
E[T_{delay}] &= \sum\limits_{k=1}^{\infty} (p_{k}\prod\nolimits_{i=1}^{k-1} (1-p_{i}))(k -\frac{1}{2}) T_{dc} \\
&= \frac{p_{1} T_{dc}}{2} + \sum\limits_{k=2}^{\infty} (1-p_{1})(1-p_{2})^{k-2} p_{2} (k -\frac{1}{2}) T_{dc} 
\end{align}

Here $p_{i}$ is mainly a function of the network load, duty cycle and resource allocation method. In contention-based protocol, we see that packets have a very low chance to arrive the destination if they do not arrive in first duty cycle. When the network load is affordable for the network, $p_{1}$ is quite high and $\{p_{i}\}^{\infty}_{i=2}$ are all similarly very low. If packets do not arrive within first duty cycle, the delay time starts to be arbitrarily large. That is the drawback of \textsc{Csma}. In schedule-based protocol, if packets do not arrive in first duty cycle, they still have a good chance to arrive in the coming cycles. In a word, we can get the following equations to calculate the information delay from the results:

\begin{align*}
E[T_{delay.contention}]&= T_{dc} (\frac{1-p_{1}}{p_{2}} + \frac{1}{2}) \\
E[T_{delay.schedule}]&= T_{dc}(\frac{1}{p_{1}} - \frac{1}{2}) \;\;\;\;\; \mbox{for} \; p_{1} = p_{2}
\end{align*}

\subsubsection{Impact of traffic intensity and burstiness}

\begin{figure}[!t]
\begin{minipage}{.49\linewidth}
  \centering
  \includegraphics[width=\linewidth]{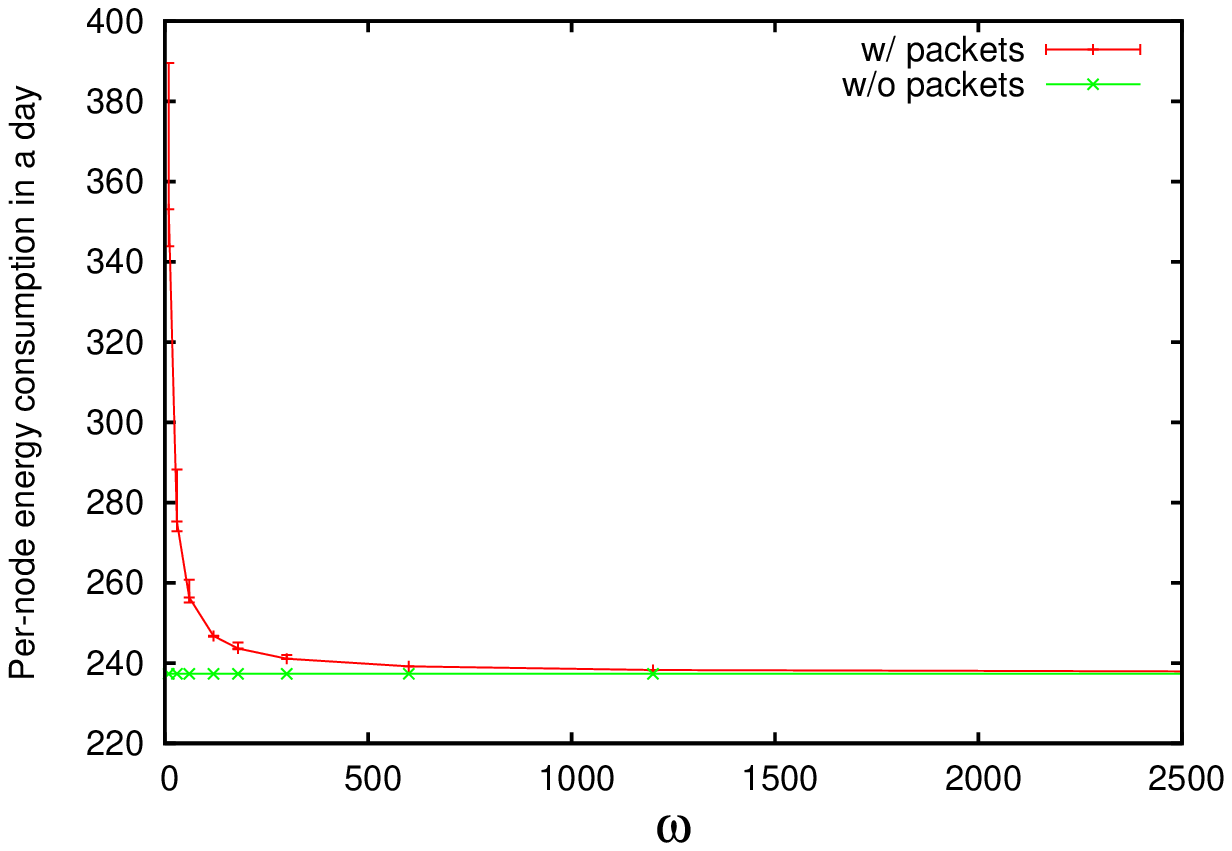}
\end{minipage}
\hfill
\begin{minipage}{.49\linewidth}
 \centering
 \includegraphics[width=\linewidth]{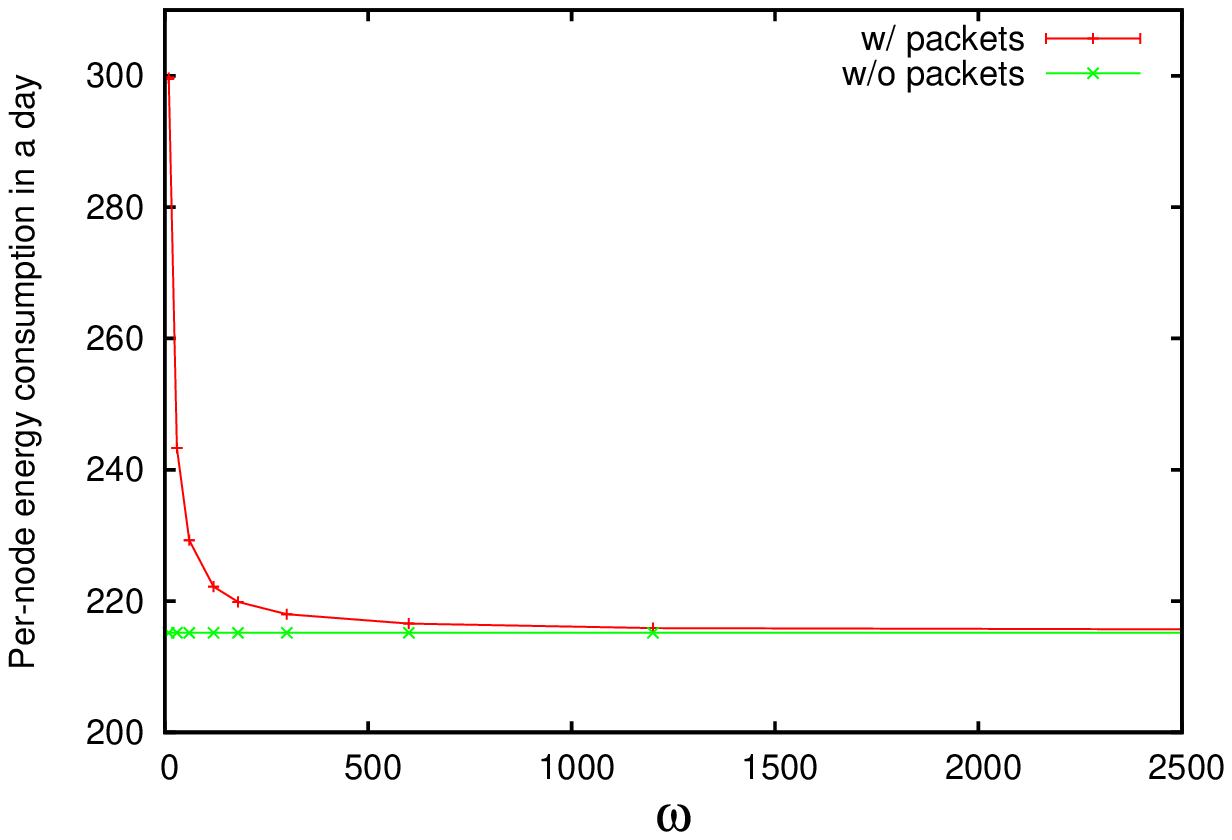}
\end{minipage}
\caption{\footnotesize{Per-node energy consumption: \emph{Periodic} application under (left) contention- and (right) schedule-based B.A. while $N=24$}}
\label{fig:time-802.15.4-tdma-pernode-energy-dynamic}
\begin{minipage}{.49\linewidth}
 \centering
 \includegraphics[width=\linewidth]{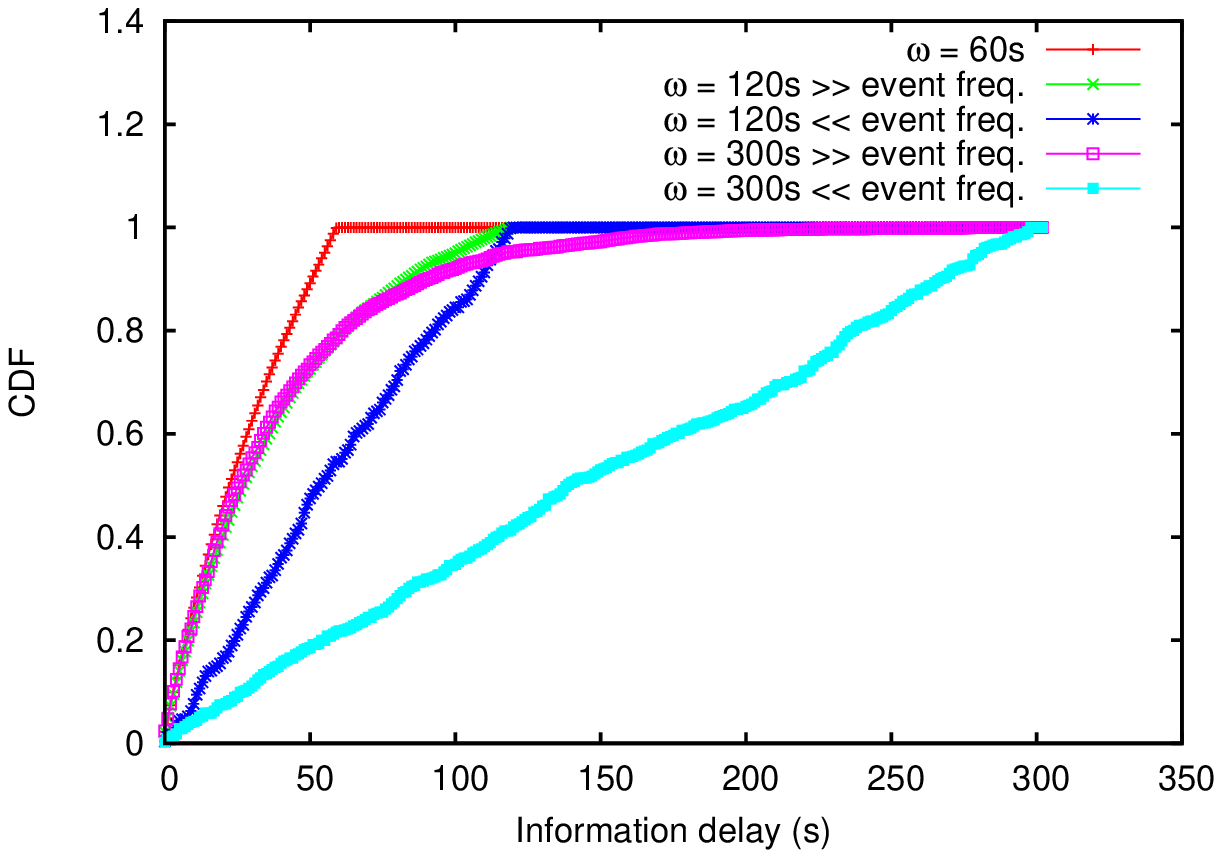}
\end{minipage}
\hfill
\begin{minipage}{.49\linewidth}
 \centering
 \includegraphics[width=\linewidth]{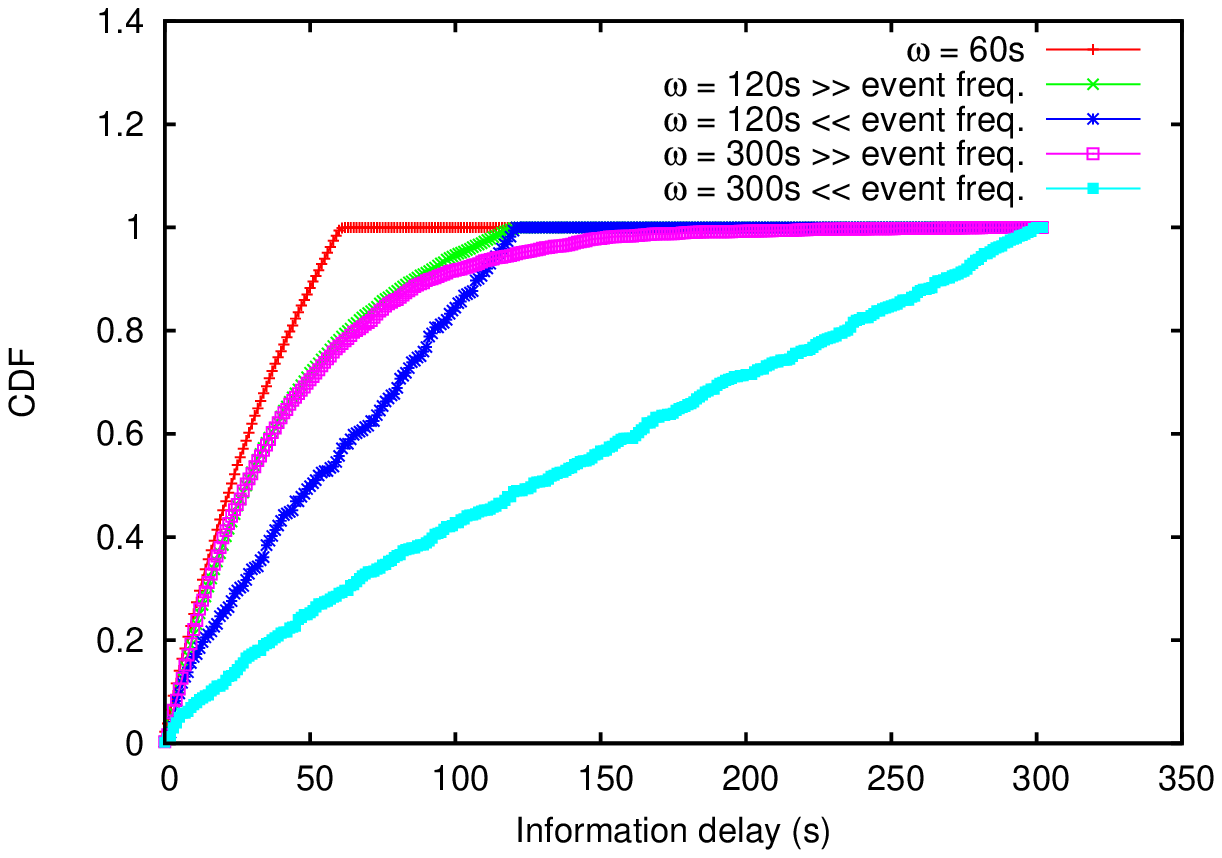}
\end{minipage}
\caption{\footnotesize{Information delay: \emph{Periodic} application while varying traffic parameters ($N=24$) under (left) contention- and (right) schedule-based B.A.}}
\label{fig:info-delay-tdma-time-event}   
\begin{minipage}{.49\linewidth}
  \centering
  \includegraphics[width=\linewidth]{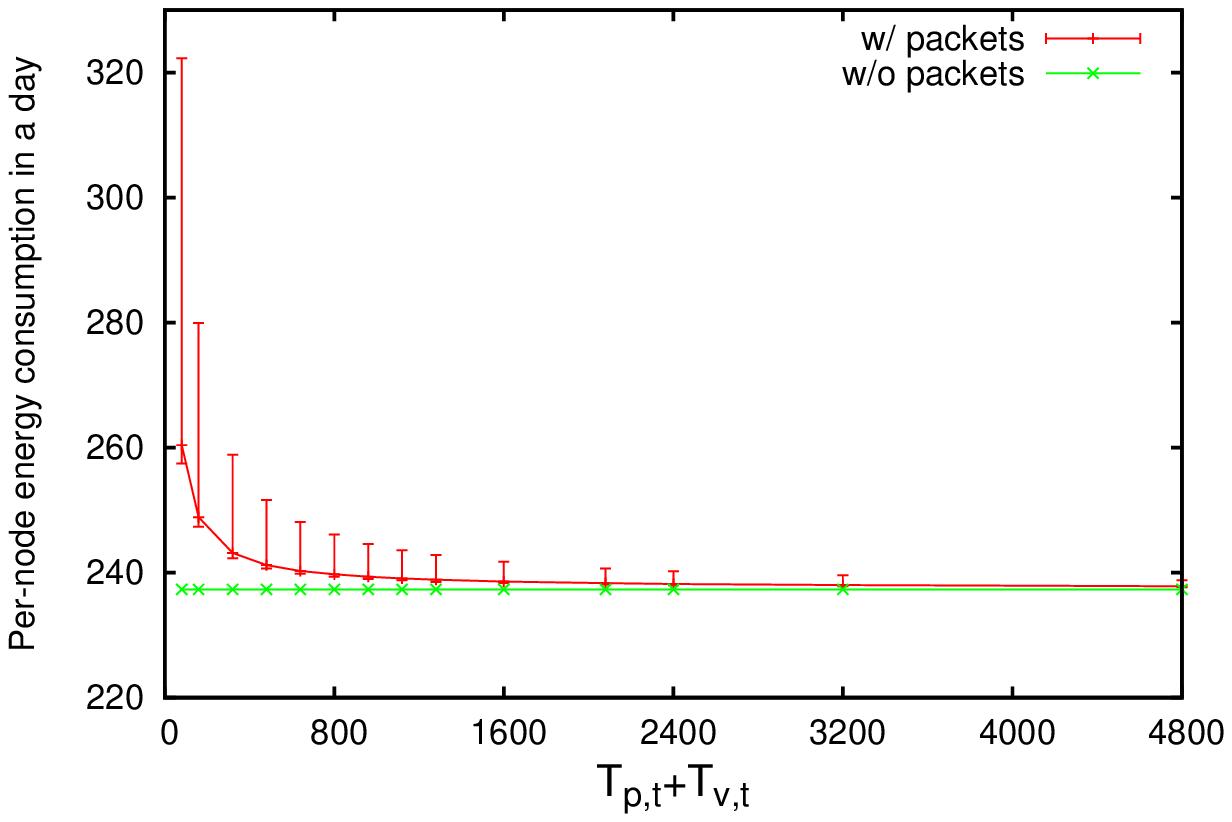}
\end{minipage}
\hfill
\begin{minipage}{.49\linewidth}
 \centering
 \includegraphics[width=\linewidth]{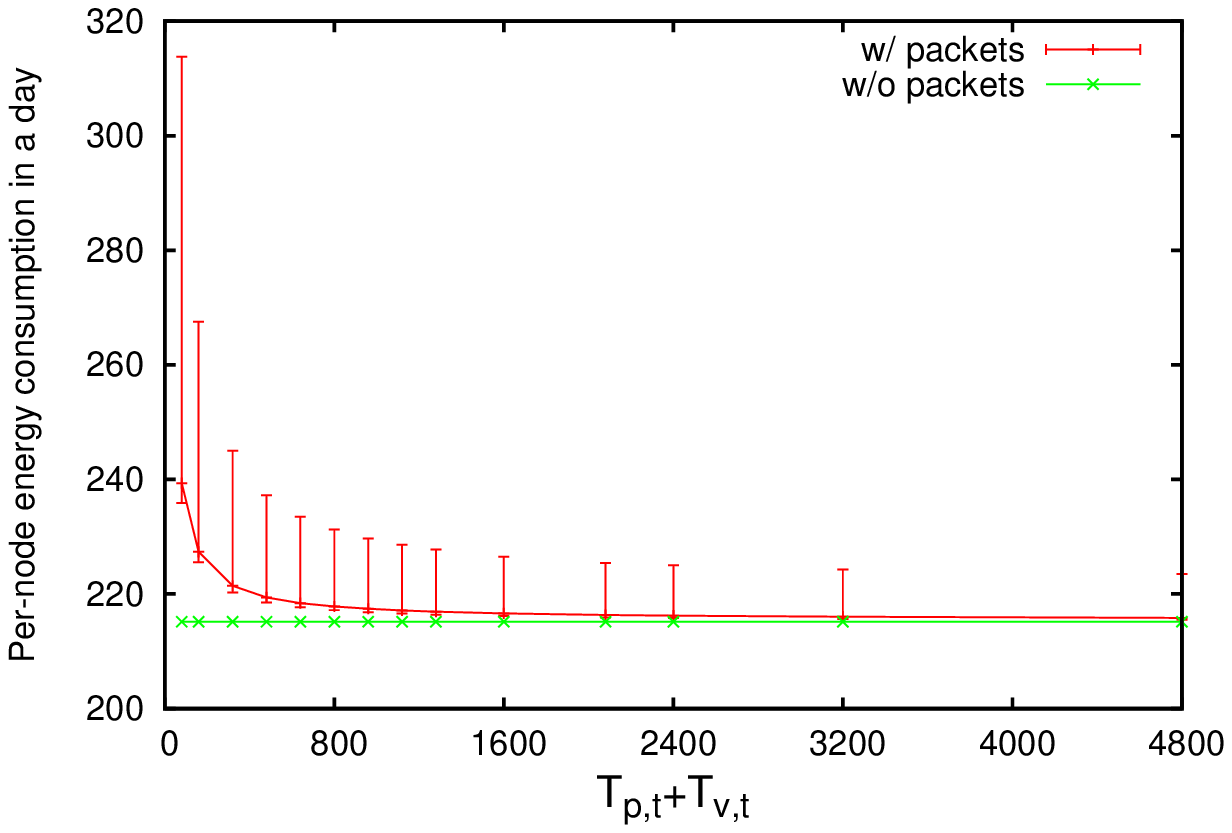}
\end{minipage}
\caption{\footnotesize{Per-node energy consumption: \emph{Event-driven} application under (left) contention- and (right) schedule-based B.A. while $N=24$}}
\label{fig:event-802.15.4-tdma-pernode-energy}
\begin{minipage}{.49\linewidth}
 \centering
 \includegraphics[width=\linewidth]{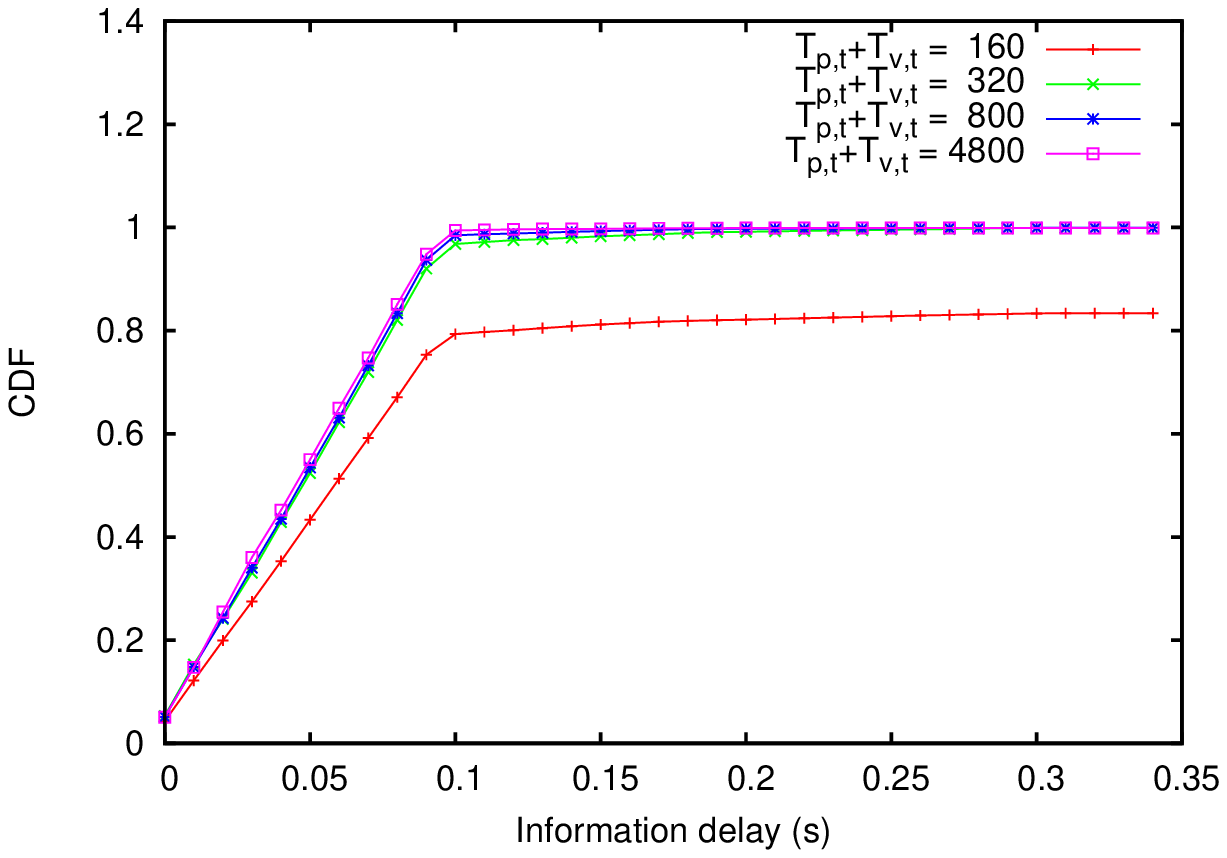}
\end{minipage}
\hfill
\begin{minipage}{.49\linewidth}
 \centering
 \includegraphics[width=\linewidth]{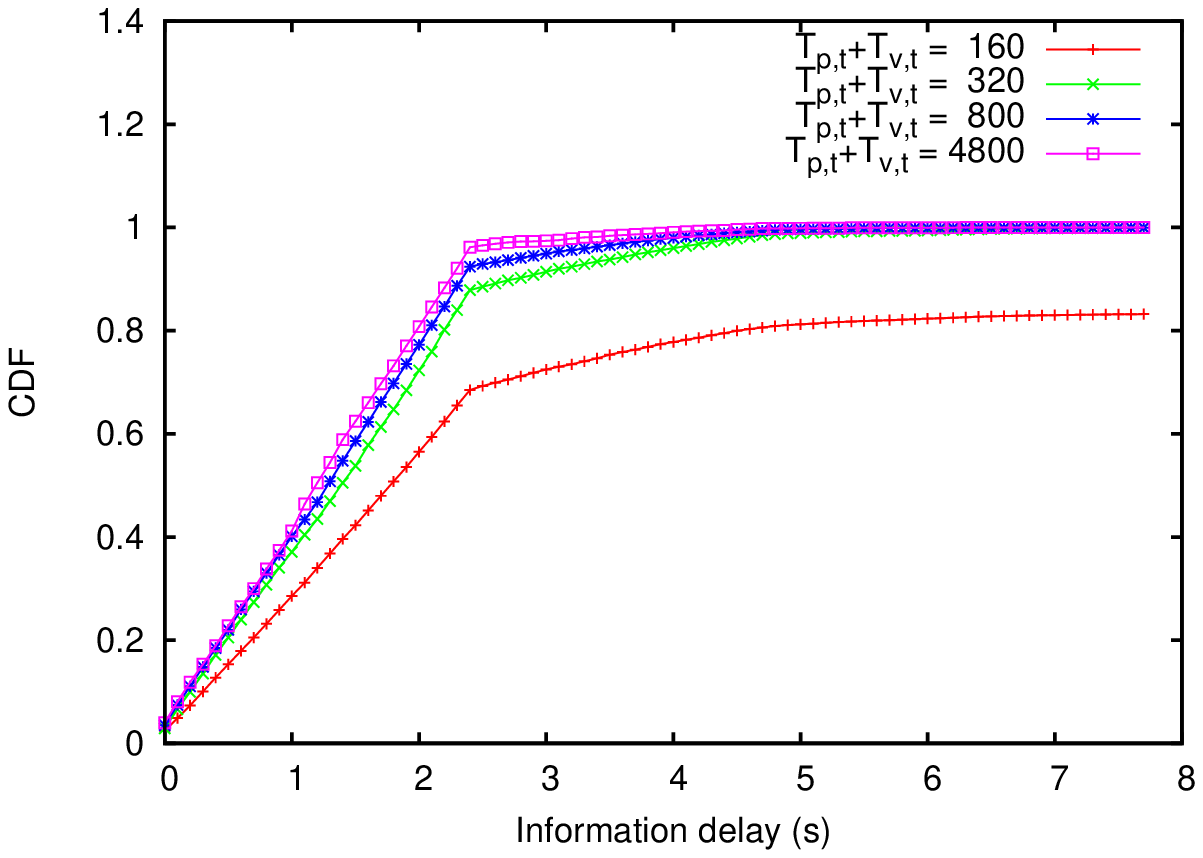}
\end{minipage}
\caption{\footnotesize{Information delay: \emph{Event-driven} application while varying traffic parameters ($N=24$) under (left) contention- and (right) schedule-based B.A.}}
\label{fig:info-delay-802.15.4-time-event}
\end{figure}

The simulation we ran in this section used the topology depicted in Figure~\ref{fig:topology}(b) with 24 nodes. Figure~\ref{fig:time-802.15.4-tdma-pernode-energy-dynamic} shows the relationship between traffic interval $\omega$ and per-node energy consumption, we see that periodic traffic is less affected by the B.A. method and strongly related to the traffic interval. The periodic traffic is equivalent to constant bit rate so that the traffic is known and uniform among all sensor nodes. Hence the deviation of consumed energy is not apparent. In Figure~\ref{fig:info-delay-tdma-time-event}, we see that the information delay between periodic and event-driven applications is not that different. Since the arriving time point during the traffic interval is uniformly distributed, its expected value can be calculated by the summation of the application-layer delay and \textsc{Mac}-layer end-to-end delay, that is, $\frac{\omega}{2} + T_{dc} (\frac{1-p_{1}}{p_{2}} + \frac{1}{2})$. If the event occurrence frequency is much higher than traffic interval, the sensory status could change more than one time in $\omega$ seconds. In this way, some update information will be missed if the sensor does not store it into packet and the information delay will be also shorter and exponentially distributed. However, it is obvious that the consumed energy is extremely low while $\omega \geq 1200$. In other words, the information delay which is more than $1200$ seconds will not be acceptable for realtime urban services. But it is interesting to assign a periodic traffic with a long interval on sensor nodes simply to inform gateways of their existences and current battery status. On the contrary, event-driven application, affected by Weibull distribution, has a larger variation on packet generation rate, and so does the energy consumption, shown in Figure~\ref{fig:event-802.15.4-tdma-pernode-energy}. The burstiness of Weibull traffic model creates a strong non-uniform traffic on each sensor node because of its infinite variance. Figure~\ref{fig:info-delay-802.15.4-time-event} shows the information delay while the network dimension is fixed. When the network load is affordable for the network ($p_{1} \rightarrow 1$), the information delay is generally proportional to $T_{dc}$. When we increase the event occurrence rate ($T_{p,t}+T{v,t} = 4800 \rightarrow 320$), $p1$ starts to reduce from 0.96 to 0.88 in schedule-based B.A. and from 0.99 to 0.96 in contention-based B.A.. Even the network load is high, contention-based protocol can still delivery more packets ($79\%$) than schedule-based ($68\%$) in first duty cycle thanks to its dynamic bandwidth allocation method. Schedule-based protocol requires more central information exchange between sensors and network coordinator so that it is much less adaptive to the traffic variation and not preferable for distant sensor nodes. 

\newpage

\subsubsection{Impact of duty cycle}

\begin{figure}[!t]
\begin{minipage}{.49\linewidth}
 \centering
 \includegraphics[width=\linewidth]{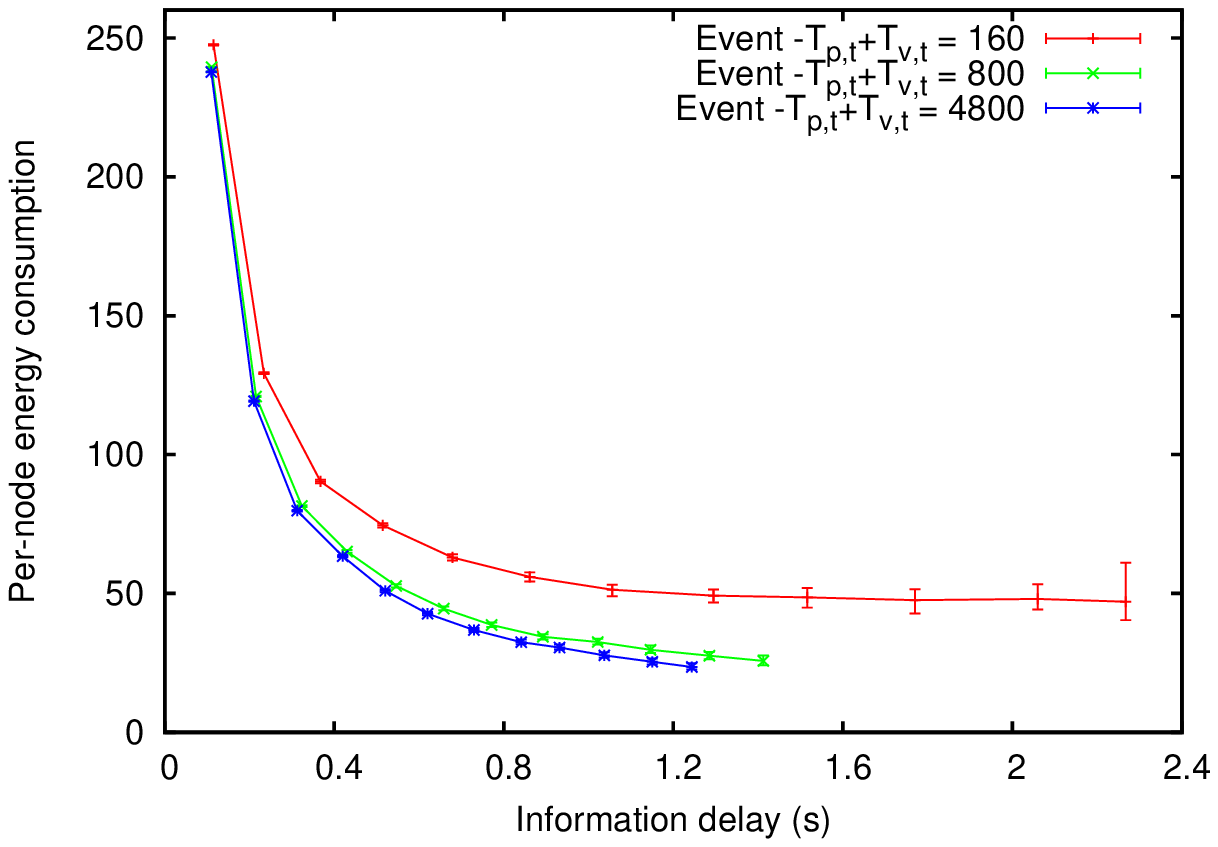}
\end{minipage}
\hfill
\begin{minipage}{.49\linewidth}
 \centering
 \includegraphics[width=\linewidth]{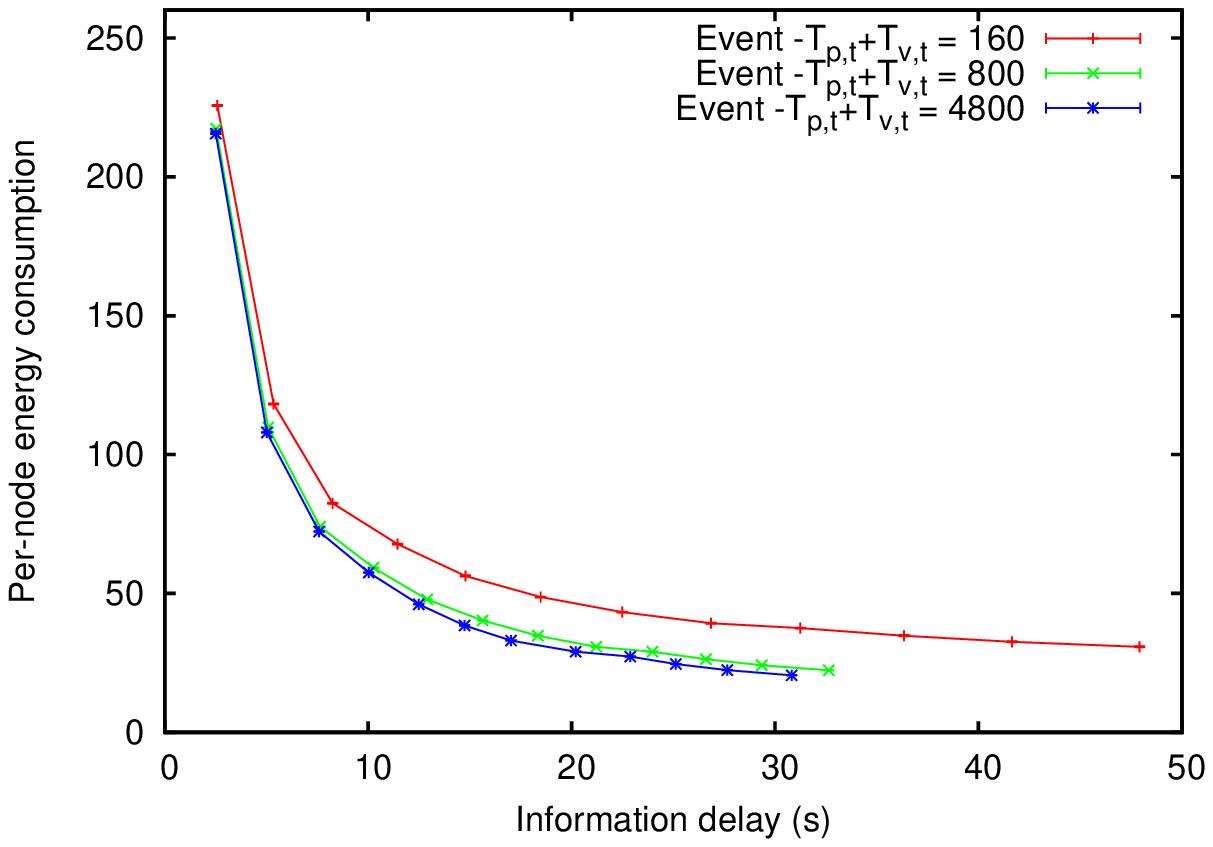}
\end{minipage}
\caption{\footnotesize{Per-node energy-delay tradeoff of event-driven application in (left) contention- and (right) schedule-based B.A. while $N=24$}}
\label{fig:802.15.4-tdma-energy-delay-tradeoff} 
\end{figure}

From Figure~\ref{fig:802.15.4-tdma-energy-delay-tradeoff}, duty cycle is generally determined by slot duration which is, however, bounded by traffic model. The minimum slot duration $T_{slot.min}$ shall be long enough to complete a reservation and a piggybacked packet transmission so that packet size and data rate are the important factors. Nevertheless, the maximum slot duration $T_{slot.max}$ is limited by the minimum required throughput according to applications carried out in the network. That is to say, if the slot duration is equal to $T_{slot}$ seconds, the maximum throughput will not exceed $\frac{1}{T_{slot}}$ packets per second by assuming that the inactive period is zero. We varied slot duration from 0.1 to 1.2 seconds by applying the topology in Figure~\ref{fig:topology}(b) with 24 nodes and then got the energy-delay tradeoff. For $T_{p,t}+T_{v,t}=160$, $T_{slot.max}<<3.33$ considering the routing overhead. Hence, we can see that the energy deviation is more obvious when the slot duration is approximating to $T_{slot.max}$ in contention-based B.A.. That is because the number of data slot decreases with $\frac{1}{T_{slot}}$ and results in more network competitors in each data slot. However, thanks to the extremely low collision rate in schedule-based B.A., the energy deviation is still low even though we increase the slot duration. Accordingly, while having the same slot duration, the energy consumption in contention-based B.A. is a bit higher, yet the information delay is much shorter. In other words, subject to the throughput conditions, for the same information delay, contention-based B.A. consumes significantly less energy than schedule-based one.

\subsubsection{Impact of multiple-hop}
\label{subsec:multi-hop}

\begin{figure}[!t]
\centering
\includegraphics[width=0.9\linewidth]{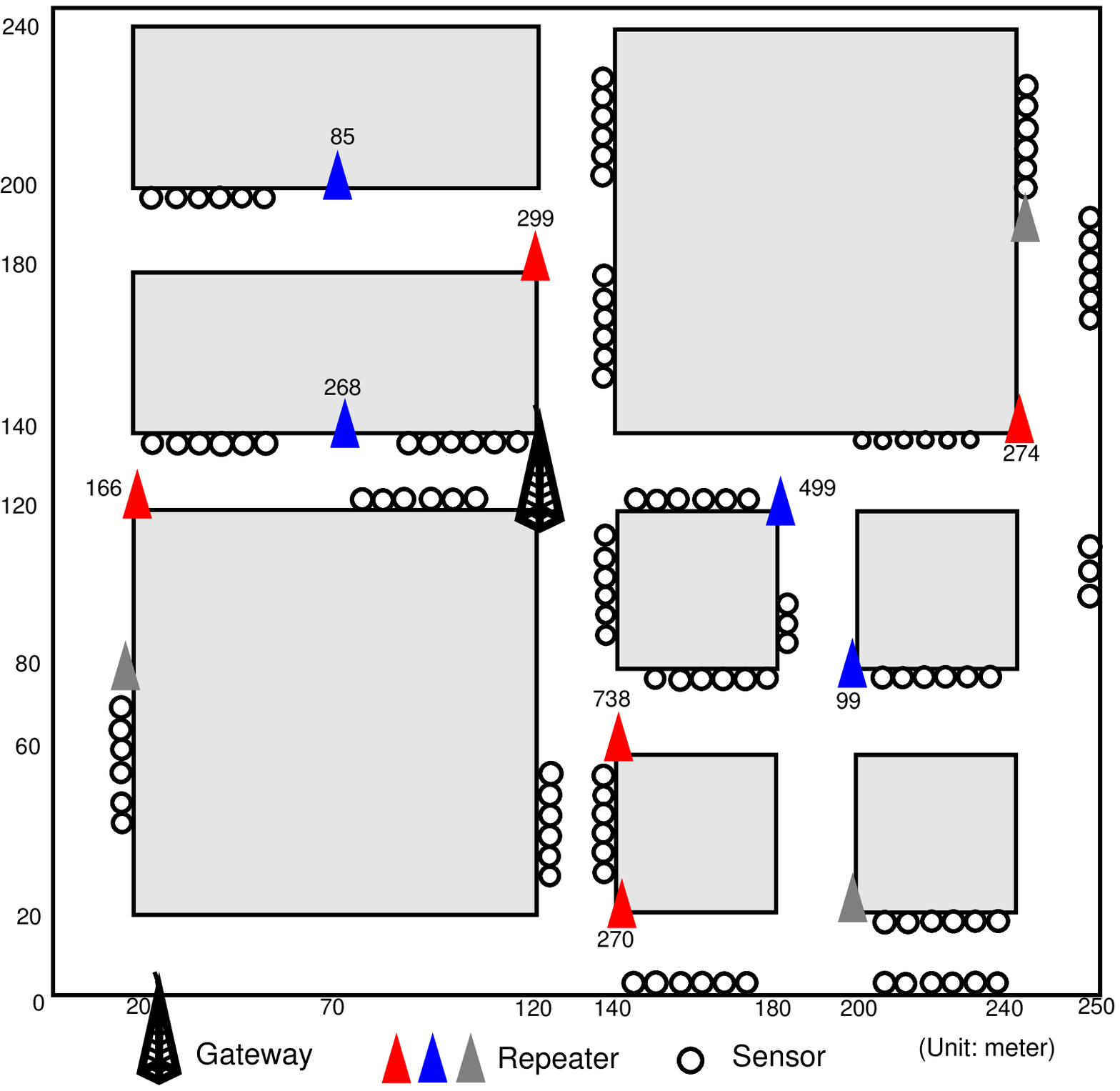}
\caption{\footnotesize{Parking Map with 120 parking sensors}}
\label{fig:mini-parking-map}
\begin{minipage}{.49\linewidth}
 \centering
 \includegraphics[width=\linewidth]{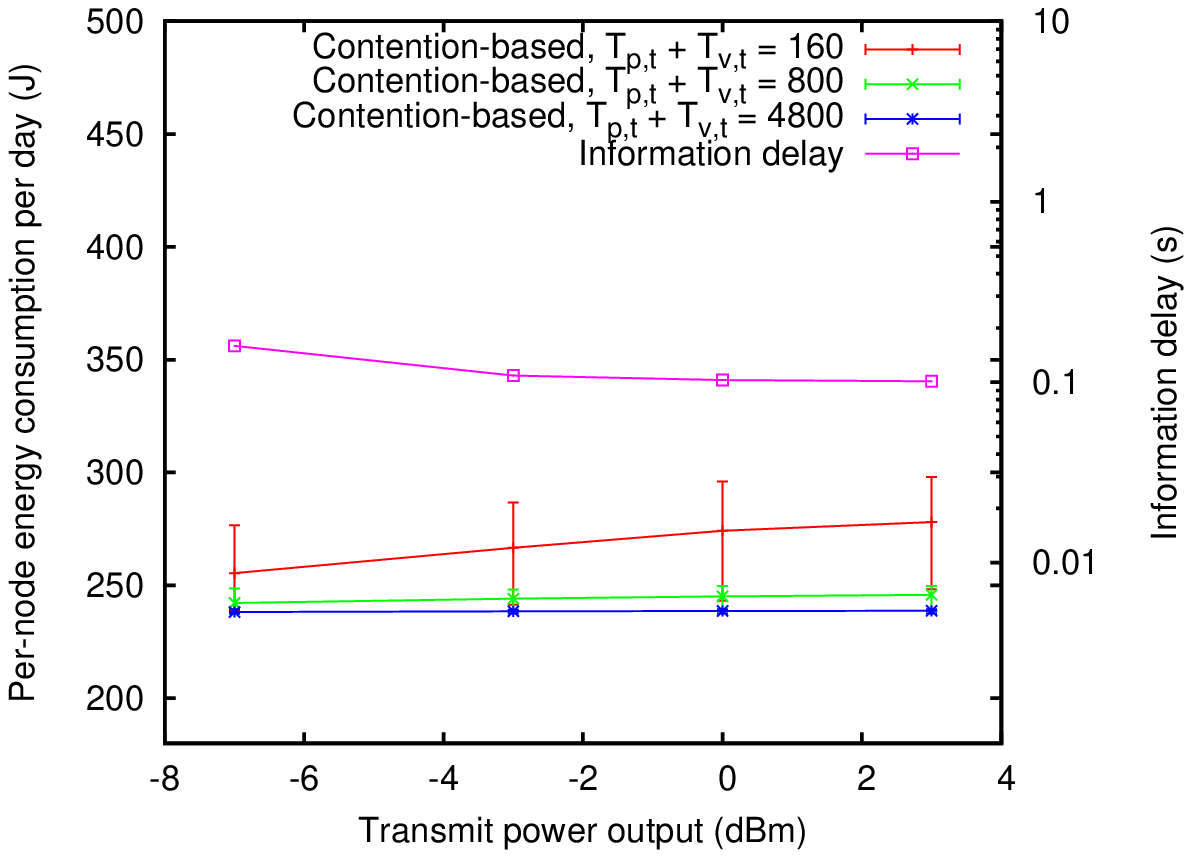}
\end{minipage}
\hfill
\begin{minipage}{.49\linewidth}
 \centering
 \includegraphics[width=\linewidth]{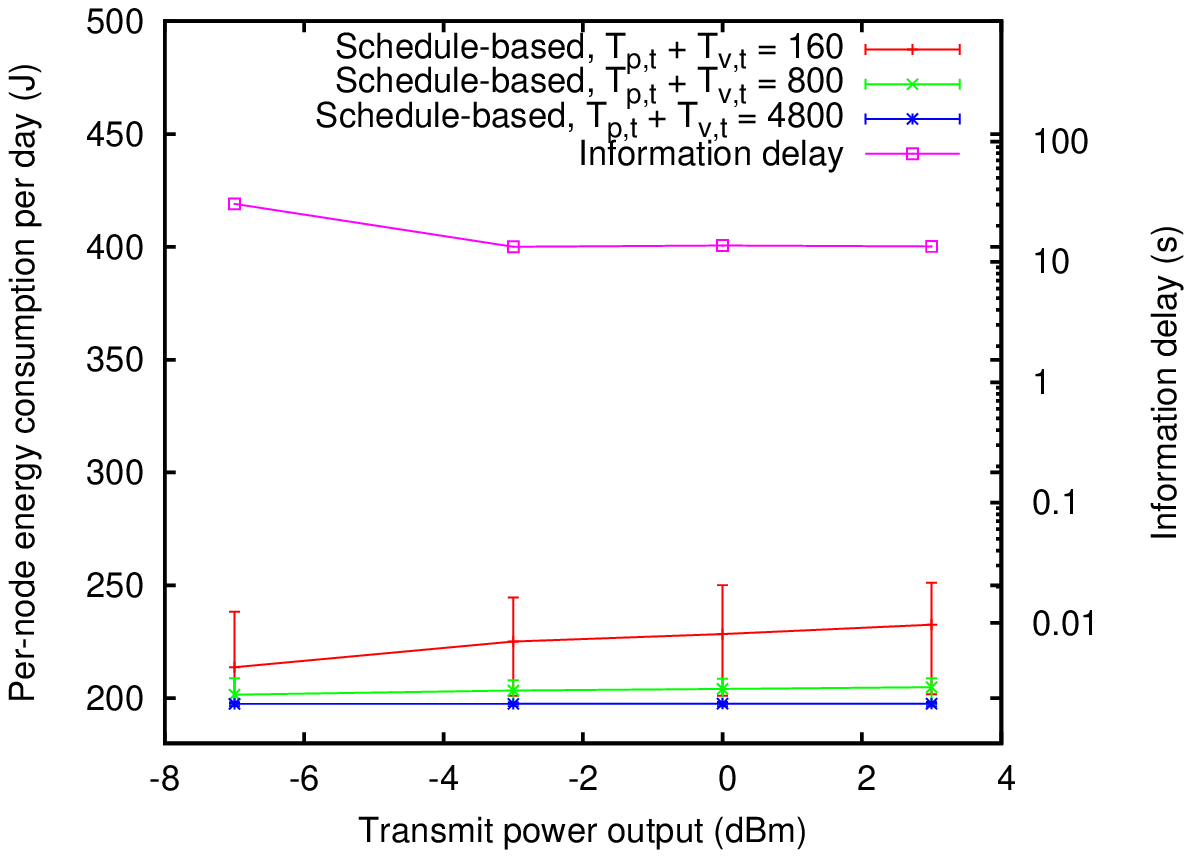}
\end{minipage}
\caption{\footnotesize{Per-node consumed energy and delay of event-driven application under (left) contention- and (right) schedule-based B.A.}}
\label{fig:802.15.4-tdma-ED}
\end{figure}

While deploying a large-scale parking sensor network, multiple hop is inevitable considering the complexity of urban environment and the limitation of transmission range. Since the transmission range is generally proportional to the transmit power output (\textsc{Tpo}), to reduce \textsc{Tpo} and transmit packets with a shorter transmission distance will be a good idea to extend the lifetime of parking sensors. The longer lifetime the sensor has, the fewer maintenance the municipality has. From the previously discussed, we see the problem of energy hole and hop limit (delay). For a better consideration, we take a parking map from SFpark\cite{SFparkreport} and construct it in our simulation depicted in Figure~\ref{fig:mini-parking-map}. Here, all sensors work as a reduced function device and can only communicate with repeaters or gateways. Only repeaters and gateway, namely full function device, can forward any received packets until the packet arrives the gateway. Since parking sensor is generally in-ground and repeater can be installed on the streetlight and powered by solar cell, this configuration is greatly preferable to municipalities. With the parking map we have, we first deploy repeaters in each intersection and in the middle of the road section when needed. In this simulation, we take four different \textsc{Tpo} = $\{3, 0, -3, -7\}$ dBm. In Figure~\ref{fig:mini-parking-map}, the red repeaters are needed when \textsc{Tpo} is equal to 3 and 0 dBm. When \textsc{Tpo} is -3 dBm, the red, blue and grey repeaters are all needed as well. When \textsc{Tpo} is -7 dBm, the grey repeater cannot provide a stable support and have to add some more repeaters to guarantee the quality of service. When \textsc{Tpo} is smaller than -10 dBm, the density of repeater will be very high (at least 2 repeaters per road section). The energy and information delay results are shown in Figure~\ref{fig:802.15.4-tdma-ED} while applying contention- and schedule-based B.A.. Also, we get some facts from this experiences: First, the update frequency of gradient routing, which affects the transmission path, shall be high enough according to the dynamic urban environment. Second, the network is divided into several small dimension cells so that contention-based protocol shall be favorable. Third, a minimum \textsc{Tpo} is required considering the router deployment limited by the street layout. Fourth, the load of each router, labeled as a numeric in Figure~\ref{fig:mini-parking-map}, shows that the load balancing is needed for gradient routing; otherwise, certain routers will run out of energy soon.

\section{Engineering Insights}
\label{sec:EngineeringInsights}

\begin{figure}[!t]
\centering
\includegraphics[width=\linewidth]{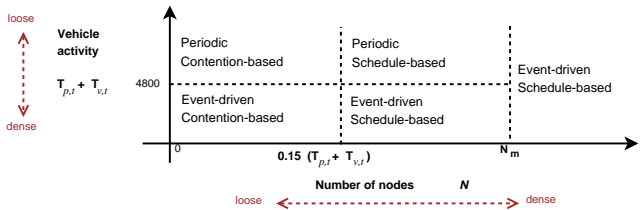}
\caption{\footnotesize{Best configuration versus vehicle activity and network density}}
\label{fig:TableProtocolSelection}
\end{figure}

In this section, we summarize our results in section~\ref{sec:experience} and provide engineering insights to streamline the \textsc{Wsn} construction of urban smart parking application. The B.A. method is the utmost important key point of determining energy consumption and information delay when traffic and node densities are known a priori. We applied two fundamental types of B.A. to our simulations instead of choosing particular protocols. In this way, we can see clearly that how the traffic intensity and node density affect the network performance and the results could serve as guidelines for urban sensor network designers. We discuss it separately from the following viewpoints by referring to Figure~\ref{fig:TableProtocolSelection}. 

\begin{itemize}
	\item \textsc{Network traffic models} We showed that the impact of the periodic traffic interval for selecting an appropriate $\omega > 1200$, and then the influence of event triggered application with Weibull distributed vehicle occupancy time. From literature, the shape parameter is always smaller than 1 and heavy-tailed. Weibull distributed packet interarrival time also shows the burstiness and non-uniform packet generation rate of network traffic comparing to exponentially distributed. The traffic characteristics requires a dynamic bandwidth allocation, e.g., contention-based \textsc{Mac}. When the event occurrence frequency is 2 times more than $\omega = 1200$, part of event-driven packets can be combined with the periodic traffic so as to reduce the network load and energy consumption after evaluating the information delay. 

	\item \textsc{Information delay} From the results, the delay time can be calculated by equation(\ref{eq:infodelay-event}) and then mainly determined by the probability of arriving in $1^{st}$ duty cycle ($p_{1}$) and the duration of duty cycle ($T_{dc}$). In schedule-based B.A., if packets do not arrive in $1^{st}$ duty cycle, they can still have quite high chance to arrive in $2^{nd}$ or $3^{rd}$ duty cycle. In contention-based B.A., the packets, which do not arrive in $1^{st}$ duty cycle, will probably never arrive or have a very long information delay due to the exponential backoff time. Thus, while the ratio of network dimension to network load is higher than $0.3$, schedule-based B.A. shall be applied to ensure the packet delivery ratio. $N = 0.3 * \frac{1}{2} * (T_{p,t} + T_{v,t}) = 0.3 ( \frac{1}{2}\lambda\gamma(1+\frac{1}{\alpha}) + \frac{1}{2}\mu\gamma(1+\frac{1}{\beta}))$.

	\item \textsc{Energy-delay tradeoff} The energy consumption is proportional to network dimension, traffic variation and duty cycle. The duty cycle is generally determined by slot duration and limited by $T_{slot.max} < \frac{1}{N}(\frac{1}{2}\lambda\gamma(1+\frac{1}{\alpha}) + \frac{1}{2}\mu\gamma(1+\frac{1}{\beta}))$. While giving a desired information delay, contention-based \textsc{Mac} consumes much less energy. What will happen if the network traffic and node densities are both high? $T_{dc}$ shall be greater than $\frac{2}{T_{p,t}+T_{v,t}}$ or $\frac{1}{\omega}$ respectively. Since $\frac{2}{T_{p,t}+T_{v,t}} \ll \frac{1}{2\omega}$ to apply periodic application, certainly $\frac{1}{\omega}$ exceeds $T_{dc}$ faster than $\frac{1}{2\omega}$. By assuming $N_{m}$ is the maximum number of nodes to apply periodic application in schedule-based B.A., we have $((N+1)*T_{slot}+T_{inactive})^{-1}=\frac{1}{\omega}$. Thus, $N_{m} = \frac{\omega}{T_{slot}}-1$ if $T_{inactive}=0$.   
	
	\item \textsc{Transmit power output} The \textsc{Psn} architecture introduced in subsection~\ref{subsec:multi-hop} is good for extending the lifetime of in-ground sensor nodes. Based on this, the communication range is significant for router/gateway deployment. A minimum \textsc{Tpo} is required for sensor nodes so as to reach these routers or gateways in crossroads.  
	
	\item \textsc{Load balancing} In multi-hop network, the routing path is maintained by gradient routing. From the results in Figure~\ref{fig:mini-parking-map}, we see the network load in each router is highly non-uniform. The load balancing shall be considered into the implementation of routing protocol. 
	
\end{itemize}

\section{Conclusion}
\label{sec:conclusion}
In this paper, we studied parking sensor networks, especially focusing on delay constraints and energy efficiency issues from a viewpoint of network traffic. Two types of traffic models are performed with different rate parameters. While deploying in different topologies and traffic variation, the network limit is revealed. With the simulation results, we then provide our engineering insights for urban sensor network designers, in particular the best combination of traffic models and bandwidth allocation depending on the urban activity and the network density. We also highlight some issues while deploying parking sensors in multi-hop networks.  

The protocol selection figure, shown in previous section, gives a good picture of the impact from the variation of traffic and node density. Each combination performs a set of information in respective circumstances. While knowing the network limits, the performance is easier to be maintained by adjusting the protocol behaviors and parameters to reach the optimization goal. Even though these thresholds can be slightly shifted by particular optimized protocols, no doubt it retains a clear overview to build urban applications over \textsc{Psn}. These insights highlight the importance to develop an adaptive \textsc{Mac} protocol which is able to distributedly detect the intensity of traffic and switch between event- and periodic traffic model when required. 








\section{Acknowledgments}
Funding for this project was provided by two grants from the Rh\^one-Alpes Region, France: F. Le Mou\"el currently holds a mobility grant Explora'Pro and T. Lin a doctoral fellowship ARC 07 n$^{o}$7075.

{\small
\bibliography{reference}

\begin{thebibliography}{10}

\bibitem{ctcontention2005}
Z.~Abichar and J.~Chang.
\newblock Conti: Constant-time contention resolution for wlan access.
\newblock In {\em NETWORKING 2005. Networking Technologies, Services, and
  Protocols; Performance of Computer and Communication Networks; Mobile and
  Wireless Communications Systems}, volume 3462 of {\em Lecture Notes in
  Computer Science}, pages 358--369. Springer Berlin Heidelberg, 2005.

\bibitem{Ahn:2006:FLS:1182807.1182837}
G.-S. Ahn, S.~G. Hong, E.~Miluzzo, A.~T. Campbell, and F.~Cuomo.
\newblock Funneling-mac: A localized, sink-oriented mac for boosting fidelity
  in sensor networks.
\newblock In {\em Proceedings of the 4th International Conference on Embedded
  Networked Sensor Systems}, SenSys '06, pages 293--306, New York, NY, USA,
  2006. ACM.

\bibitem{6662948}
M.~Arfeen, K.~Pawlikowski, D.~McNickle, and A.~Willig.
\newblock The role of the weibull distribution in internet traffic modeling.
\newblock In {\em Teletraffic Congress (ITC), 2013 25th International}, pages
  1--8, Sept 2013.

\bibitem{EricBradlow}
E.~Bradlow, P.~Fader, M.~Adrian, and B.~B. McShane.
\newblock Count models based on weibull interarrival times.
\newblock {\em Journal of Business \& Ecnomic Statistics}, 26, 2008.

\bibitem{926982}
W.~Heinzelman, A.~Chandrakasan, and H.~Balakrishnan.
\newblock Energy-efficient communication protocol for wireless microsensor
  networks.
\newblock In {\em System Sciences, 2000. Proceedings of the 33rd Annual Hawaii
  International Conference on}, pages 10 pp. vol.2--, Jan 2000.

\bibitem{jamieson2006sift}
K.~Jamieson, H.~Balakrishnan, and Y.~Tay.
\newblock {Sift: a MAC Protocol for Event-Driven Wireless Sensor Networks}.
\newblock In {\em Third European Workshop on Wireless Sensor Networks (EWSN)},
  Zurich, Switzerland, February 2006.

\bibitem{4622710}
J.~Kim, J.~On, S.~Kim, and J.~Lee.
\newblock Performance evaluation of synchronous and asynchronous mac protocols
  for wireless sensor networks.
\newblock In {\em Sensor Technologies and Applications, 2008. SENSORCOMM '08.
  Second International Conference on}, pages 500--506, Aug 2008.

\bibitem{thesisdequentin}
Q.~Lampin.
\newblock {\em Wireless Urban Sensor network: Applications, characterization
  and protocols (in French)}.
\newblock {PhD} dissertation, INSA de Lyon, 2014.

\bibitem{6214027}
Q.~Lampin, D.~Barthel, I.~Auge-Blum, and F.~Valois.
\newblock Cascading tournament mac: Low power, high capacity medium sharing for
  wireless sensor networks.
\newblock In {\em Wireless Communications and Networking Conference (WCNC),
  2012 IEEE}, pages 1544--1549, 2012.

\bibitem{Mathur:2010:PDS:1814433.1814448}
S.~Mathur, T.~Jin, N.~Kasturirangan, J.~Chandrasekaran, W.~Xue, M.~Gruteser,
  and W.~Trappe.
\newblock Parknet: drive-by sensing of road-side parking statistics.
\newblock In {\em Proceedings of the 8th international conference on Mobile
  systems, applications, and services}, MobiSys '10, pages 123--136, New York,
  NY, USA, 2010. ACM.

\bibitem{Mitov2006555}
K.~V. Mitov and N.~M. Yanev.
\newblock Superposition of renewal processes with heavy-tailed interarrival
  times.
\newblock {\em Statistics \& Probability Letters}, 76(6):555 -- 561, 2006.

\bibitem{Pister08tsmp:time}
K.~S.~J. Pister and L.~Doherty.
\newblock Tsmp: Time synchronized mesh protocol.
\newblock In {\em In Proceedings of the IASTED International Symposium on
  Distributed Sensor Networks (DSN08)}, 2008.

\bibitem{6583499}
E.~Polycarpou, L.~Lambrinos, and E.~Protopapadakis.
\newblock Smart parking solutions for urban areas.
\newblock In {\em 2013 IEEE 14th International Symposium and Workshops on a
  World of Wireless Mobile and Multimedia Networks (WoWMoM)}, pages 1--6, 2013.

\bibitem{Rajendran:2003:ECM:958491.958513}
V.~Rajendran, K.~Obraczka, and J.~J. Garcia-Luna-Aceves.
\newblock Energy-efficient collision-free medium access control for wireless
  sensor networks.
\newblock In {\em Proceedings of the 1st International Conference on Embedded
  Networked Sensor Systems}, SenSys '03, pages 181--192, New York, NY, USA,
  2003. ACM.

\bibitem{6179195}
G.~Revathi and V.~Dhulipala.
\newblock Smart parking systems and sensors: A survey.
\newblock In {\em 2012 International Conference on Computing, Communication and
  Applications (ICCCA)}, pages 1--5, 2012.

\bibitem{4453818}
I.~Rhee, A.~Warrier, M.~Aia, J.~Min, and M.~Sichitiu.
\newblock Z-mac: A hybrid mac for wireless sensor networks.
\newblock {\em Networking, IEEE/ACM Transactions on}, 16(3):511--524, June
  2008.

\bibitem{Rhee:2006:DDR:1132905.1132927}
I.~Rhee, A.~Warrier, J.~Min, and L.~Xu.
\newblock Drand: Distributed randomized tdma scheduling for wireless ad-hoc
  networks.
\newblock In {\em Proceedings of the 7th ACM International Symposium on Mobile
  Ad Hoc Networking and Computing}, MobiHoc '06, pages 190--201, New York, NY,
  USA, 2006. ACM.

\bibitem{UrbanMobility2012}
D.~Schrank, B.~Eisele, , and L.~T.
\newblock {\em Urban Mobility Report 2012}.
\newblock Texas Transportation Institute, The Texas A\& M University System,
  Texas USA, 2012.

\bibitem{SFparkreport}
SFMTA.
\newblock Sfpark: Putting theory into practive.
\newblock Available at
  \url{http://sfpark.org/wp-content/uploads/2011/09/sfpark_aug2011projsummary_print-2.pdf},
  August 2011.

\bibitem{6488838}
D.~Stanislowski, X.~Vilajosana, Q.~Wang, T.~Watteyne, and K.~Pister.
\newblock Adaptive synchronization in ieee802.15.4e networks.
\newblock {\em IEEE Transactions on Industrial Informatics}, PP(99):1--1, 2013.

\bibitem{1492678}
Q.~Sun, S.~Tan, and K.~Teh.
\newblock Analytical formulae for path loss prediction in urban street grid
  microcellular environments.
\newblock {\em IEEE Transactions on Vehicular Technology}, 54(4):1251--1258,
  July 2005.

\bibitem{Sun:2008:RRA:1460412.1460414}
Y.~Sun, O.~Gurewitz, and D.~B. Johnson.
\newblock Ri-mac: A receiver-initiated asynchronous duty cycle mac protocol for
  dynamic traffic loads in wireless sensor networks.
\newblock In {\em Proceedings of the 6th ACM Conference on Embedded Network
  Sensor Systems}, SenSys '08, pages 1--14, New York, NY, USA, 2008. ACM.

\bibitem{SmartSantanderReport}
E.~I. Vlahogianni, K.~Kepaptsoglou, V.~Tsetsos, and M.~G. Karlaftis.
\newblock Exploiting new sensor technologies for real-time parking prediction
  in urban areas.
\newblock In {\em Transportation Research Board 93rd Annual Meeting Compendium
  of Papers}, 14-1673, jan 2014.

\bibitem{Watteyne-5425543}
T.~Watteyne, K.~Pister, D.~Barthel, M.~Dohler, and I.~Auge-Blum.
\newblock Implementation of gradient routing in wireless sensor networks.
\newblock In {\em Global Telecommunications Conference, 2009. GLOBECOM 2009.
  IEEE}, pages 1--6, 2009.

\bibitem{WSNet}
An event-driven simulator for large scale wireless sensor networks.
\newblock Available at \url{http://wsnet.gforge.inria.fr}.

\bibitem{Ye:2006:UDC:1182807.1182839}
W.~Ye, F.~Silva, and J.~Heidemann.
\newblock Ultra-low duty cycle mac with scheduled channel polling.
\newblock In {\em Proceedings of the 4th International Conference on Embedded
  Networked Sensor Systems}, SenSys '06, pages 321--334. ACM, 2006.

\bibitem{6644967}
S.~Zhuo, Z.~Wang, Y.-Q. Song, Z.~Wang, and L.~Almeida.
\newblock iqueue-mac: A traffic adaptive duty-cycled mac protocol with dynamic
  slot allocation.
\newblock In {\em Sensor, Mesh and Ad Hoc Communications and Networks (SECON),
  2013 10th Annual IEEE Communications Society Conference on}, pages 95--103,
  June 2013.

\end{thebibliography}
\bibliographystyle{abbrv}
}


%


\end{document}